\documentclass[useAMS,usegraphicx,usenatbib]{mn2e}

\usepackage{amssymb}
\usepackage{color}

\usepackage{epstopdf}
\usepackage{amsmath}

\newcommand\lsim{\mathrel{\rlap{\lower4pt\hbox{\hskip1pt$\sim$}}
\raise1pt\hbox{$<$}}}
\newcommand\gsim{\mathrel{\rlap{\lower4pt\hbox{\hskip1pt$\sim$}}
\raise1pt\hbox{$>$}}} 
 
 \newcommand{\tot}{\mathrm{tot}}

\newcommand{\Ham}{\mathcal{H}}

\newcommand{\itot}{i_\textnormal{tot}}

\newcommand{\msun}{M_\odot}

\title[Secular Dynamics in Hierarchical Three-Body Systems] {Secular Dynamics in Hierarchical Three-Body Systems }

\author[Naoz et al.]{Smadar Naoz$^{1,2,\dagger}$, Will M.\ Farr$^{1}$,
  Yoram Lithwick$^{1,3}$, Frederic A.\
  Rasio$^{1,3}$,  Jean Teyssandier$^{1,4}$ 
\\ $^1$CIERA, Northwestern University, Evanston, IL 60208,
  USA \\ $^2$ Institute for Theory and Computation, Harvard-Smithsonian Center for Astrophysics, 60 Garden St.; Cambridge, MA, USA 02138 \\ $^3$ Department of Physics and Astronomy, Northwestern
  University \\
  $^4$ Institut d'Astrophysique de Paris, UMR 7095, CNRS, UPMC, 98 bis bd Arago, F-75014 Paris\\
  $\dagger$ Einstein Fellow
}
\begin{document}

\label{firstpage}

\maketitle

\begin{abstract}
The secular approximation for the evolution of hierarchical triple
  configurations has proven to be very useful in many astrophysical
  contexts, from planetary to triple-star systems.  In this approximation the orbits
  may change shape and orientation, on time scales longer than the orbital time scales,  but the semimajor axes are
  constant.  For example, for highly inclined triple systems, the
  Kozai-Lidov mechanism can produce large-amplitude oscillations of
  the eccentricities and inclinations.  Here we revisit the secular
  dynamics of hierarchical triple systems.  We derive the secular
  evolution equations to octupole order in Hamiltonian perturbation
  theory.  Our derivation corrects an error in some previous
  treatments of the problem that implicitly assumed a conservation of the  z-component of the angular momentum of the inner orbit (i.e., parallel to the total angular momentum of the system).  
   Already to quadrupole order, our results show new behaviors including the possibility for a system to  oscillate from prograde to retrograde orbits.
 At the octupole order, for an eccentric outer orbit, the inner orbit can reach extremely high eccentricities and
  undergo chaotic flips in its orientation.  We discuss applications
  to a variety of astrophysical systems, from stellar triples to
  merging compact binaries and planetary systems.  Our results agree
  with those of previous studies done to quadrupole order only in the
  limit in which one of the inner two bodies is a massless test
  particle and the outer orbit is circular; 
  our results agree with previous studies at octupole order for the eccentricity evolution, but not for the inclination evolution.
%  our results agree with
%  previous studies at octupole order in the eccentricity evolution,
%  but the inclination evolution we present has not been considered by
%  previous studies.
\end{abstract}

\section{Introduction}\label{intro}

Triple star systems are believed to be very common
\citep[e.g.,][]{T97,Eggleton+07}. From dynamical stability arguments
these must be hierarchical triples, in which the (inner) binary is
orbited by a third body on a much wider orbit.  Probably more than
50\% of bright stars are at least double
\citep{T97,Eggleton+07}. Given the selection effects against finding
faint and distant companions we can be reasonably confident that the
proportion is actually substantially greater.  \citet{T97} showed that
$40\%$ of binary stars with period $<10$~d in which the primary is a
dwarf ($0.5 -1.5\,M_{\odot}$) have at least one additional
companion. He found that the fraction of triples and higher multiples
among binaries with period ($10-100\,$d) is $\sim10\%$. Moreover,
\citet{Pri+06} have surveyed a sample of contact binaries, and noted
that among 151 contact binaries brighter than 10 mag., 42$\pm5\%$ are
at least triple.

Many close stellar binaries with two compact objects are likely
produced through triple evolution.  Secular effects (i.e., coherent
interactions on timescales long compared to the orbital period), and
specifically Kozai-Lidov cycling \citep[][see below]{Kozai,Lidov},
have been proposed as an important element in the evolution of triple
stars \citep[e.g.][]{Har69,Mazeh+79,Sod82,1998KEM,Dan,PF09,Tho10,ST12}.  In
addition, Kozai-Lidov cycling has been suggested to play an important
role in both the growth of black holes at the centers of dense star
clusters and the formation of short-period binary black holes
\citep{Wen,MH02,Bla+02}.  Recently, \citet{Iva+10} showed that the
most important formation mechanism for black hole XRBs in globular
clusters may be triple-induced mass transfer in a black hole-white
dwarf binary.

Secular perturbations in triple systems also play an important role in
planetary system dynamics.  \citet{Kozai} studied the effects of
Jupiter's gravitational perturbation on an inclined asteroid in our
own solar system. In the assumed hierarchical configuration, treating
the asteroid as a test particle, \citet{Kozai} found that its
inclination and eccentricity fluctuate on timescales much larger than
its orbital period. Jupiter, assumed to be in a circular orbit,
carries most of the angular momentum of the system. Due to Jupiter's
circular orbit and the negligible mass of the asteroid, the system's
potential is axisymmetric and thus the component of the inner orbit's
angular momentum along the total angular momentum is conserved during
the evolution.  \citet{Kozia79} also showed the importance of secular
interactions for the dynamics of comets \citep[see
also][]{Quinn+90,Bailey+92,Thomas+96}.  The evolution of the orbits of
binary minor planets is dominated by the secular gravitational
perturbation from the sun \citep{PN09}; properly accounting for the
resulting secular effects---including Kozai cycling---accurately
reproduces the binary minor planet orbital distribution seen today
\citep{Naoz10obs,Grundy+11}. In addition \citet{Kin+91},
\citet{vas99}, \cite{Carr+02}, \citet{Nes+03}, \citet{cuk+04} and
\citet{KN07} suggested that secular interactions may explain the
significant inclinations of gas giant satellites and Jovian irregular
satellites.

Similar analyses have been applied to the orbits of extrasolar planets
\citep[e.g.,][]{Inn+97,Wu+03,Dan,Wu+07,Naoz10,Veras+10,Cor+11}. \citet{Naoz10}
considered the secular evolution of a triple system consisting of an
inner binary containing a star and a Jupiter-like planet at several
AU, orbited by a distant Jupiter-like planet or brown-dwarf
companion. Perturbations from the outer body can drive Kozai-like
cycles in the inner binary, which, when planet-star tidal effects are
incorporated, can lead to the capture of the inner planet onto a
close, highly-inclined or even retrograde orbit, similar to the orbits
of the observed retrograde ``hot Jupiters.''  Many other studies of
exoplanet dynamics have considered similar systems, but with a very
distant stellar binary companion acting as perturber. In such systems,
the outer star completely dominates the orbital angular momentum, and
the problem reduces to test-particle evolution \citep[see
][]{LN,Boaz2,Naoz+12}. If the lowest level of approximation is applicable
(e.g., the outer perturber is on a circular orbit), the $z$-component
of the inner orbit's angular momentum is conserved
\citep[e.g.,][]{LZ74}.

In early studies of high-inclination secular perturbations
\citep{Kozai,Lidov}, the outer orbit was circular and again dominated
the orbital angular momentum of the system. In this situation, the
component of the inner orbit's angular momentum along the z-axis is
conserved.  In many later studies the assumption that the
$z$-component of the inner orbit's angular momentum is constant was
built into the equations
\citep[e.g.][]{1998EKH,Mik+98,Zdz+07}. %,Takeda}. 
In fact these studies
are only valid in the limit of a test particle forced by a perturber
on a circular orbit.  To leading order in the ratio of semimajor axes,
the double averaged potential of the outer orbit is axisymmetric (even
for an eccentric outer perturber), thus if taken to the test particle
limit, this results in a conservation of the $z$-component of the
inner orbit's angular momentum.  We refer to this limit as the
``standard'' treatment of Kozai oscillations,
i.e. quadrupole-level approximation in the test particle limit (test
particle quadrupole, hereafter TPQ).

In this paper we show that a common mistake in the Hamiltoniano
treatment of these secular systems can lead to the erroneous
conclusion that the $z$-component of the inner orbit's angular
momentum is constant outside the TPQ limit; in fact, the $z$-component
of the inner orbit's angular momentum is only conserved by the
evolution in the test-particle limit and to quadrupole order.
To demonstrate the error we focus on the quadrupole
(non-test-particle) approximation in the main body of the paper, but
we include the full octupole--order equations of motion in an appendix.  

{In what follows we show the applications of these two effects (i.e., correcting the error and including the full octupole--order equations of motion) by considering different astrophysical systems. Note that the applications illustrated in the text are inspired by real systems; however, we caution that we consider here only Newtonian point mass dynamics, while in reality other effects such as tides and general relativity can greatly effect the evolution. For example, general relativity may alter the evolution of the system, which can give rise to a resonant behavior of the inner orbitÕs eccentricity  \citep[e.g.,][]{Ford00,Naoz12b}. Furthermore, tidal forces can suppress the eccentricity growth of the inner orbit, and thus significantly modify the evolutionary track of the system  \citep[e.g.,][]{Mazeh+79,Sod84,1998KEM}. In particular tides, in some cases, can considerably suppress the chaotic behavior that arises in the presence of the the octupoleÐlevel of approximation  \citep[e.g.,][]{Naoz10,Naoz+12}. Therefore, while the examples presented in this paper are inspired by real astronomical systems,
the true evolutionary behavior will be modified from what we show
once the eccentricity
becomes too high. }

This paper is organized as follows. We first present the general
framework (\S \ref{Sec:Ham}); we then derive the complete formalism
for the quadrupole-level approximation and the equations of motion (\S
\ref{sec:4eq}), we also develop the octupole-level approximation
equations of motion in \S \ref{sec:H8}. We discuss a few of the most
important implications of the correct formalism in \S
\ref{sec:implications}. We also compare our results with those of
previous studies (\S \ref{sec:implications}) and offer some
conclusions in \S \ref{Sec:con}.

\section{Hamiltonian Perturbation Theory for Hierarchical Triple
  Systems}\label{Sec:Ham}

Many gravitational triple systems are in a hierarchical
configuration---two objects orbit each other in a relatively tight
inner binary while the third object is on a much wider orbit.  If the
third object is sufficiently distant, an analytic, perturbative
approach can be used to calculate the evolution of the system.  In the
usual secular approximation \citep[e.g.,][]{Maechal90},
 the two orbits torque each other and exchange angular momentum, but
not energy.  Therefore the orbits can change shape and orientation (on
timescales much longer than their orbital periods), but not semimajor
axes (SMA).
\begin{figure}
\begin{center}
%\plotone{config1.eps}
\includegraphics[width=84mm]{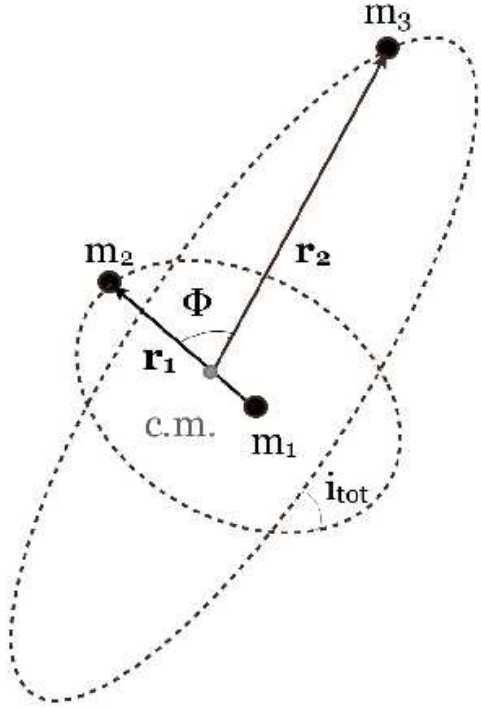}
\caption{Coordinate system used to describe the hierarchical triple
  system ({\it not to scale}). Here 'c.m.' denotes the center of mass
  of the inner binary, containing objects of masses $m_1$ and
  $m_2$. The separation vector ${\bf r}_1$ points from $m_1$ to $m_2$;
  ${\bf r}_2$ points from 'c.m.' to $m_3$. The angle between the
  vectors ${\bf r}_1$ and ${\bf r}_2$ is $\Phi$. } \label{fig:config}
\end{center}
\end{figure}

We first define our basic notations. The system consists of a close
binary (bodies of masses $m_1$ and $m_2$) and a third body (mass
$m_3$).  It is convenient to describe the orbits using Jacobi
coordinates \citep[][p. 441-443]{MD00}. Let ${\bf r}_1$ be the
relative position vector from $m_1$ to $m_2$ and ${\bf r}_2$ the
position vector of $m_3$ relative to the center of mass of the inner
binary (see fig.~\ref{fig:config}).  Using this coordinate system the
dominant motion of the triple can be divided into two separate
Keplerian orbits: the relative orbit of bodies~1 and~2, and the orbit
of body~3 around the center of mass of bodies~1 and~2. The Hamiltonian
for the system can be decomposed accordingly into two Keplerian
Hamiltonians plus a coupling term that describes the (weak)
interaction between the two orbits.  Let the SMAs of the inner and
outer orbits be $a_1$ and $a_2$, respectively. Then the coupling term
in the complete Hamiltonian can be written as a power series in the
ratio of the semi-major axes $\alpha=a_1/a_2$
\citep[e.g.,][]{Har68}. In a hierarchical system, by definition, this
parameter $\alpha$ is small.

The complete Hamiltonian expanded in orders of $\alpha$ is
\citep[e.g.,][]{Har68},
\begin{eqnarray}
\label{eq:Ham}
\Ham&=&\frac{k^2m_1m_2}{2a_1}+\frac{k^2 m_3(m_1+m_2)}{2a_2} \\ \nonumber
&& + \frac{k^2}{a_2} \sum_{j=2}^\infty\alpha^j M_j \left(\frac{r_1}{a_1}\right)^j\left(\frac{a_2}{r_2}\right)^{j+1}P_j(\cos{\Phi}) \ ,
\end{eqnarray}
where $k^2$ is the gravitational constant, $P_j$ are Legendre
polynomials, $\Phi$ is the angle between ${\bf r}_1$ and ${\bf r}_2$
(see Figure \ref{fig:config}) and
\begin{equation}
\label{eq:Mj}
M_j=m_1m_2m_3\frac{m_1^{j-1}-(-m_2)^{j-1}}{(m_1+m_2)^{j}} \ .
\end{equation}
Note that we have followed the convention of \citet{Har69} and chosen
our Hamiltonian to be the negative of the total energy, so that $\Ham
> 0$ for bound systems.

\begin{figure}
\begin{center}
\includegraphics[width=84mm]{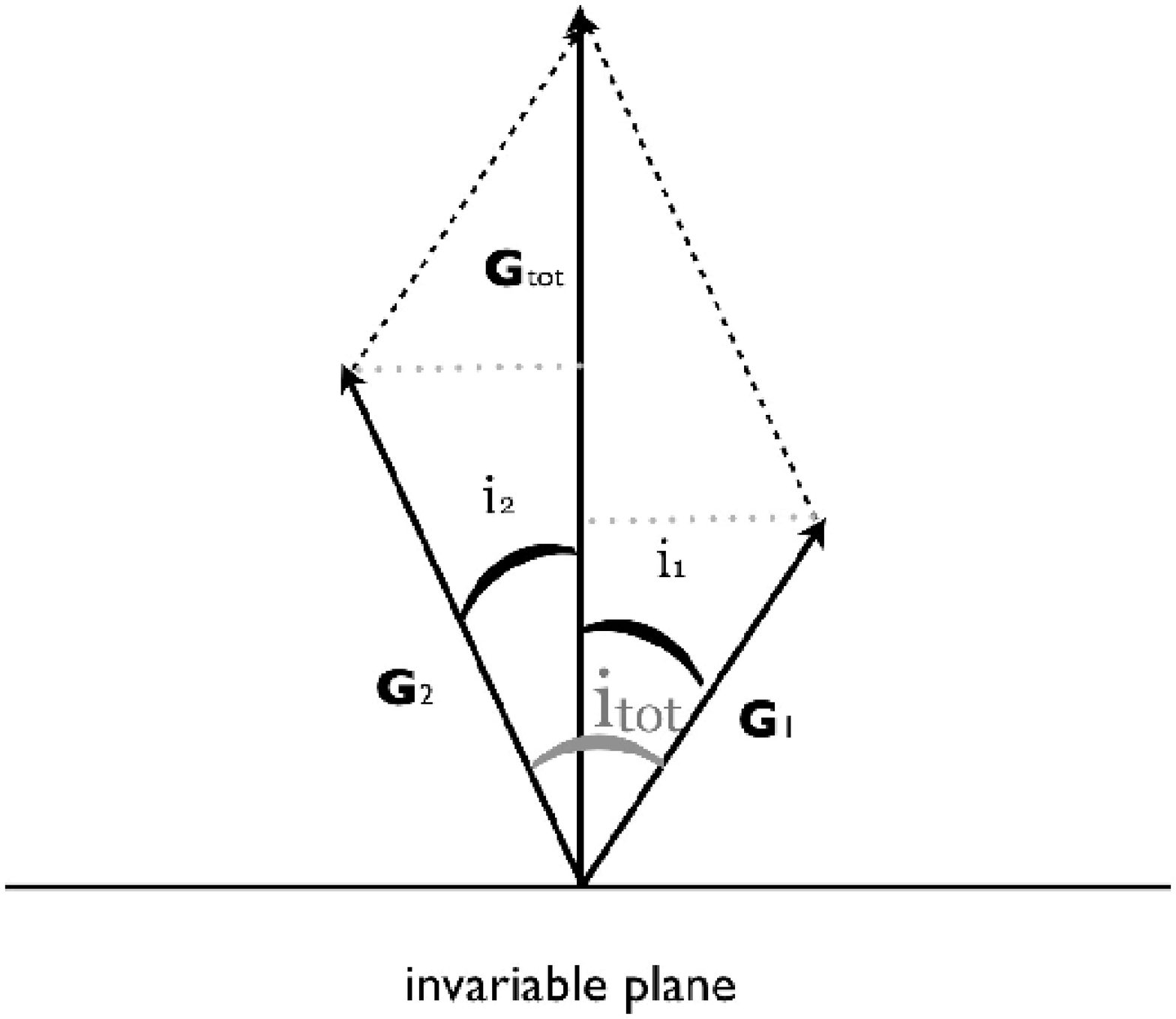}
\caption{Geometry of the angular momentum vectors. We show the total
  angular momentum vector (${\bf G}_\tot$), the angular momentum vector
  of the inner orbit (${\bf G}_1$) with inclination $i_1$ with respect
  to ${\bf G}_\tot$ and the angular momentum vector of the outer orbit
  (${\bf G}_2$) with inclination $i_2$ with respect to
  ${\bf G}_\tot$. The angle between ${\bf G}_1$ and ${\bf G}_2$ defines
  the mutual inclination $i_\tot=i_1+i_2$. The invariable plane is
  perpendicular to ${\bf G}_\tot$.} \label{fig:angular}
\end{center}
\end{figure}

We adopt the canonical variables known as Delaunay's elements, which
provide a particularly convenient dynamical description of our
three-body system \citep[e.g.][]{3book}. The coordinates are chosen to
be the mean anomalies, $l_1$ and $l_2$, the longitudes of ascending
nodes, $h_1$ and $h_2$, and the arguments of periastron, $g_1$ and
$g_2$, where subscripts $1,\,2$ denote the inner and outer orbits,
respectively. Their conjugate momenta are
\begin{eqnarray}
L_1&=&\frac{m_1 m_2}{m_1+m_2}\sqrt{k^2(m_1+m_2)a_1} \ , \label{eq:L1}\\ \nonumber
L_2&=&\frac{m_3(m_1+ m_2)}{m_1+m_2+m_3}\sqrt{k^2(m_1+m_2+m_3)a_2} \ , \label{eq:L2}
\end{eqnarray}
\begin{equation}
G_1=L_1\sqrt{1-e_1^2} \ , \quad G_2=L_2\sqrt{1-e_2^2} \ ,
\end{equation}
and
\begin{equation}
H_1=G_1\cos{i_1} \ , \quad H_2=G_2\cos{i_2} \  ,
\end{equation}
where $e_1$ ($e_2$) is the inner (outer) orbit eccentricity.  Note
that $G_1$ and $G_2$ are also the magnitudes of the angular momentum
vectors (${\bf G}_1$ and ${\bf G}_2$), and $H_1$ and $H_2$ are the
$z$-components of these vectors.  Figure~\ref{fig:angular} shows the
resulting configuration of theses vectors. The following geometric
relations between the momenta follow from the law of cosines:
\begin{eqnarray}
\cos{\itot}&=&\frac{G_\tot^2-G_1^2-G_2^2}{2G_1G_2} \ , \label{eq:cosi} \\
H_1&=&\frac{G_\tot^2+G_1^2-G_2^2}{2G_\tot} \ , \label{eq:H1} \\
H_2&=&\frac{G_\tot^2+G_2^2-G_1^2}{2G_\tot} \ , \label{eq:H2}
\end{eqnarray}
where ${\bf G}_{\tot}={\bf G}_1+{\bf G}_2$ is the (conserved) total
angular momentum, and the angle between $\bf{G}_1$ and $\bf{G}_2$
defines the mutual inclination $i_\tot=i_1+i_2$.  From
eqs.~(\ref{eq:H1}) and~(\ref{eq:H2}) we find that the inclinations
$i_1$ and $i_2$ are determined by the orbital angular momenta:
\begin{eqnarray}
\cos i_1&=&\frac{G_{\tot}^2+G_1^2-G_2^2}{2G_{\tot}G_1} \ , \label{eq:i1} \\
\cos i_2&=&\frac{G_{\tot}^2+G_2^2-G_1^2}{2G_{\tot}G_2} \ . \label{eq:i2}
\end{eqnarray}
In addition to these geometrical relations we also have that
\begin{equation}
\label{eq:z-sum-to-total}
H_1 + H_2 = G_{\tot} = {\rm const} \ .
\end{equation}

The canonical relations give the equations of motion:
\begin{eqnarray}
\label{eq:Canoni}
\frac{dL_j}{dt}=\frac{\partial \Ham}{\partial l_j} \ , \quad \frac{dl_j}{dt}=-\frac{\partial \Ham}{\partial L_j} \ , \\
\frac{dG_j}{dt}=\frac{\partial \Ham}{\partial g_j} \ , \quad \frac{dg_j}{dt}=-\frac{\partial \Ham}{\partial G_j} \ , \\
\frac{dH_j}{dt}=\frac{\partial \Ham}{\partial h_j} \ , \quad \frac{dh_j}{dt}=-\frac{\partial \Ham}{\partial H_j} \ ,
\label{eq:Canoni3}
\end{eqnarray}
where $j=1,2$. Note that these canonical relations have the opposite
sign relative to the usual relations \citep[e.g.,][]{Goldstein}
 because of the sign convention we have chosen for our Hamiltonian.
Finally we write the Hamiltonian through second order in $\alpha$ as
\citep[e.g., ][]{Kozai}
\begin{eqnarray}
\label{eq:Ham4}
\Ham&=&\frac{\beta_1}{2L_1^2}+\frac{\beta_2}{2L_2^2} + \\ \nonumber
&& 4\beta_3\left(\frac{L_1^4}{L_2^6}\right)\left(\frac{r_1}{a_1}\right)^2\left(\frac{a_2}{r_2}\right)^3\left(3\cos2\Phi+1\right) \ ,
\end{eqnarray}
 where the mass parameters are
\begin{eqnarray}
\label{eq:beta}
\beta_1&=& k^2m_1m_2\frac{L_1^2}{a_1} \ , \\
\beta_2&=& k^2(m_1+m_2)m_3\frac{L_2^2}{a_2} \\ \quad {\rm and} \nonumber \\
\beta_3 & = & \frac{k^4}{16} \frac{\left( m_1 + m_2 \right)^7 m_3^7}{\left( m_1 m_2 \right)^3 \left( m_1 + m_2 + m_3 \right)^3} \ .
%\beta_3&=& \frac{k^2}{16}M_2\left( \frac{a_1}{L_12}\right)2\left(\frac{L_22}{a_2}\right)3 \ ,
\end{eqnarray}

\section{Secular Evolution  to the  Quadrupole Order
}\label{sec:4eq}

In this section, we derive the secular quadrupole--level Hamiltonian.  in Appendix \ref{Sec:form} we develop the
complete quadrupole-level secular approximation and in particular  in Appendix  \ref{sec:quad_eq_motion} we present the quadrupole--level  equations of motion.  The main difference
between the derivation shown here (see also Appendix \ref{Sec:form})
and those of previous studies lies in the ``elimination of nodes''
\citep[e.g.,][]{Kozai,JM66}. This is related to the transition to a
coordinate system with the total angular momentum along the z-axis,
which is known as the \emph{invariable plane}
\citep[e.g.,][]{MD00}. In this coordinate system (see Figure
\ref{fig:angular}), the longitudes of the ascending nodes differ by
$\pi$, i.e.,
\begin{equation}
\label{eq:pi}
h_1 - h_2 = \pi \ . 
\end{equation}
Conservation of the {\it total} angular momentum implies that this relation holds at
all times.  Many previous works have exploited it to explicitly
simplify the Hamiltonian by setting $h_1 - h_2 = \pi$ before  deriving the
equations of motion.  After the substitution, the
Hamiltonian is independent of the longitudes of ascending nodes ($h_1$
and $h_2$), and this can 
 lead to the incorrect conclusion that
$\dot{H}_1 = \dot{H}_2 = 0$ when the canonical equations of motion are
derived. Some previous studies incorrectly concluded that  the
$z$-components of the orbital angular momenta are always constant (see also Appendix
\ref{prob}).  The substitution $h_1 - h_2 = \pi$ is incorrect at the
Hamiltonian level because it unduly restricts variations in the
trajectory of the system to those where $\delta h_1 = \delta h_2$.
After deriving the equations of motion, however, we can exploit the
relation $h_1 - h_2 = \Delta h = \pi$, which comes from the
conservation of angular momentum. This considerably simplifies the
evolution equations.
 We
 show (Appendices \ref{sec:quad_eq_motion} and \ref{sec:8}) that one can still use the Hamiltonian with the nodes
eliminated found in previous studies \citep[e.g.,][]{Kozai,Har69} as
long as the evolution equations for the inclinations are derived from
the total angular momentum conservation, instead of using the
canonical relations. Of course, the correct evolution equations can
also be calculated from the correct Hamiltonian (without the nodes
eliminated), which we derive in this section.
%
%Some previous studies made the substitution $h_1 - h_2 = \pi$ directly
%in the Hamiltonian (see \S \ref{prob}).  After the substitution, the
%Hamiltonian is independent of the longitudes of ascending nodes ($h_1$
%and $h_2$), and thus gives an evolution where both $H_1$ and $H_2$ are
%constant.  The substitution $h_1 - h_2 = \pi$ is incorrect at the
%Hamiltonian level because it unduly restricts variations in the
%trajectory of the system to those where $\delta h_1 = \delta h_2$.
%After deriving the equations of motion, however, we can exploit the
%relation $h_1 - h_2 = \Delta h = \pi$, which comes from the
%conservation of angular momentum. This considerably simplifies the
%evolution equations.

We note that there are some other derivations of the secular evolution
equations that avoid the elimination of the nodes
\citep{Far+10,Las+10,Mar10,Boaz}, and thus do not suffer from this
error\footnote{It is possible to eliminate the nodes as long as one
  does not conclude that the conjugate momenta are constant, one
  example is \citet{Lidov+76} and another is \citet{Malige} that after
  eliminating the nodes introduced a different transformation which
  overcame the problem.}.

The secular Hamiltonian is given by the average over the
rapidly-varying $l_1$ and $l_2$ in equation~\eqref{eq:Ham4} (see
Appendix \ref{Sec:form} for more details)
\begin{eqnarray} \label{eq:hami4}
\Ham_2&=&\frac{C_2}{8}\big \{  [1+3\cos(2i_2)] \big( [2+3e_1^2] [1+3\cos(2i_1)] \\ \nonumber 
&+&30e_1^2\cos(2g_1)\sin^2(i_1) \big) + 3\cos(2\Delta h) [10 e_1^2\cos(2g_1) \\ \nonumber
&\times&\{3+\cos(2i_1)\} + 4(2+3e_1^2)\sin(i_1)^2 ] \sin^2(i_2) \\ \nonumber
&+&12 \{ 2+3e_1^2-5e_1^2\cos(2g_1)\}\cos(\Delta h)\sin(2i_1)\sin(2i_2)  \\ \nonumber
&+&120e^2_1\sin(i_1)\sin(2i_2)\sin(2g_1)\sin(\Delta h) \\ \nonumber
&-& 120e_1^2\cos(i_1)\sin^2(i_2)\sin(2g_1)\sin(2\Delta h) \big \} \ , 
\end{eqnarray}
where 
%% C_{2,our}=4*C_{2,Ford}
\begin{equation}\label{eq:C2}
  C_2=\frac{k^4}{16}\frac{(m_1+m_2)^7}{(m_1+m_2+m_3)^3}\frac{m_3^7}{(m_1 m_2)^3}\frac{L_1^4}{L_2^3 G_2^3} \ .
\end{equation}

Making the usual (incorrect) substitution $\Delta h \to \pi$
(i.e.~eliminating the nodes), we get the quadrupole-level Hamiltonian
that has appeared in many previous works \citep[see, e.g.][]{Ford00}:
\begin{eqnarray}
  \Ham_2(\Delta h \to \pi)& =& C_2 \{ \left( 2 + 3 e_1^2 \right) \left( 3 \cos^2 i_\tot - 1 \right)  \\ \nonumber
  &+&  15 e_1^2 \sin^2 i_\tot \cos(2 g_1) \} \ ,
\end{eqnarray}
where we have set $i_1+i_2=i_\tot$. Because this Hamiltonian is
missing the longitudes of ascending nodes ($h_1$ and $h_2$), many
previous studies concluded that the $z$-components (i.e.~vertical
components) of the angular momenta of the inner and outer orbits
(i.e., $H_1$ and $H_2$) are constants.

We derive the quadrupole--level equations of motions in Appendix
\ref{sec:quad_eq_motion}.  In particular, we give the equations of
motion of the z-component of the angular momentum of the inner and
outer orbits derived from the Hamiltonian in Eq.~\eqref{eq:hami4}.  As
we show in the subsequent sections the evolution of $H_{1,2}$ produces
a qualitatively different evolutionary route for many astrophysical
systems considered in previous works.
   
In Appendix~\ref{sec:maxmin} we show that the quadrupole approximation
leads to well-defined minimum and maximum eccentricity and
inclination.  The eccentricity of the inner orbit and the inner (and
mutual) inclination oscillate. In the test-particle limit, our
formalism gives the critical initial mutual inclination angles for
large oscillations of $39.2^\circ \leq \itot \leq 140.8^\circ$ with
nearly-zero initial inner eccentricity, in agreement with
\citet{Kozai}.

It is easy to show that $H_1$ and $H_2$ are constant only in the TPQ
limit without using the explicit equations of motion in
Appendix~\ref{Sec:form}.  Because the Hamiltonian in
Eq.~\eqref{eq:hami4} is independent of $g_2$, $G_2 = \mathrm{const}$
at the quadrupole level.  Combining this with the geometric relation
in Eq.~\eqref{eq:H1}, $H_1=(G_{\rm tot}^2+G_1^2-G_2^2)/(2G_{\rm
  tot})$, and the constantcy of the total angular momentum, $G_{\rm
  tot}$, we have that
\begin{equation}
\dot{H}_1 = \frac{\dot{G}_1 G_1}{G_\mathrm{tot}}.
\end{equation}
In the TPQ limit, $G_\mathrm{tot} \gg G_1$, so $\dot{H}_1 = -\dot{H}_2
\approx 0$, and the $z$-component of each orbit's angular momentum is
conserved.  Outside this limit, when $G_1/G_\mathrm{tot}$ is not
negligable, $H_1$ and $H_2$ cannot be constant.  Note that the TPQ
limit, where $G_1 \ll G_\mathrm{tot}$, is equivalent to the limit
where $i_2 \approx 0$ appearing in many previous works.

\section{Octupole-Level Evolution}\label{sec:H8}

In Appendix~\ref{sec:8}, we derive the secular evolution equations to
octupole order.  Many previous octupole--order derivations provided
correct secular evolution equations for at least some of the elements,
in spite of using the elimination of nodes substitution at the
Hamiltonian level
\citep[e.g.][]{Har68,Har69,Sid83,KM99,Ford00,Bla+02,Lee03,Tho10}. This
is because the evolution equations for $e_2$, $g_2$, $g_1$ and $e_1$
can be found correctly from a Hamiltonian that has had $h_1$ and $h_2$
eliminated by the relation $h_1 - h_2 = \pi$; the partial derivatives
with respect to the other coordinates and momenta are not affected by
the substitution. The correct evolution of $H_1$ and $H_2$ can then be
derived, not from the canonical relations, but from \emph{total}
angular momentum conservation. We discuss in more details the
comparison between this work and previous analyses in \S
\ref{sec:implications}. 

The octupole-level terms in the Hamiltonian can become important when
the eccentricity of the outer orbit is non-zero, and if $\alpha$ is
large enough.  We quantify this by considering the ratio between the
octupole to quadrupole-level coefficients, which is
\begin{equation}
\frac{C_3}{C_2}=\frac{15}{4}\left(\frac{m_1-m_2}{m_1+m_2}\right)\left(\frac{a_1}{a_2}\right)\frac{1}{1-e_2^2} \ ,
\end{equation}
where $C_3$ is the octupole-level coefficient [eq.~(\ref{eq:C3})] and
$C_2$ is the quadrupole-level coefficient [eq.~(\ref{eq:C2})].  We
define
\begin{equation}\label{eq:epsiM}
\epsilon_M=\left(\frac{m_1-m_2}{m_1+m_2}\right)\left( \frac{a_1}{a_2}\right)\frac{e_2}{1-e_2^2} \ ,
\end{equation}
which gives the relative significance of the octupole-level term in
the Hamiltonian.  This parameter has three important parts; first the
eccentricity of the outer orbit ($e_2$), second, the mass difference
of the inner binary ($m_1$ and $m_2$) and the SMA ratio%
\footnote{Note here that the subscripts ``1'' and ``2'' refer to the
  \emph{inner} bodies in $m_1$ and $m_2$, but the subscript ``2''
  refers to the \emph{outer} body in $e_2$.}. %
In the test particle limit (i.e., $m_1\gg m_2$) $\epsilon_M$ is
reduced to the octupole coefficient introduced in \citet{LN} and
\citet{Boaz2},
\begin{equation}
\epsilon=\left(\frac{a_1}{a_2}\right)\frac{e_2}{1-e_2^2} \ .
\end{equation}
We call the octupole-level behavior of a system for which $\epsilon_M
\ll 1$ is not satisfied the ``eccentric Kozai-Lidov" (EKL) mechanism.

The octupole terms vanish when $e_2=0$.  Therefore if one artificially
held $e_2=0$, in the test-particle limit the inner body's orbit would
be given by the equations derived by \cite{Kozai}, i.e.  by the test
particle quadrupole equations. However, at octupole order the value of
$e_2$ evolves in time if the inner body is massive. Furthermore, even
if the inner body is massless, if the outer body has $e_2>0$ then the
inner body's behavior will also be different than in Kozai's
treatment.  For example, \citet{LN} and \citet{Boaz2} find that the
inner orbit can flip orientation (see below) even in the
test-particle, octupole limit.  The octupole-level effects can change
qualitatively the evolution of a system. Compared to the
quadrupole-level behavior, the eccentricity of the inner orbit in the
EKL mechanism can reach a much higher value. In some cases these
excursions to very high eccentricities are accompanied by a ``flip''
of the orbit with respect to the total angular momentum, i.e.~starting
with $i_1<90^\circ$ the inner orbit can eventually reach
$i_1>90^\circ$ (see Figures \ref{fig:hypoplanet3}--\ref{fig:hypostar}
for examples).  Chaotic behavior is also possible at the octupole
level \citep{LN}, but not at the quadrupole--level.

\begin{figure}
\begin{center}
\includegraphics[width=84mm]{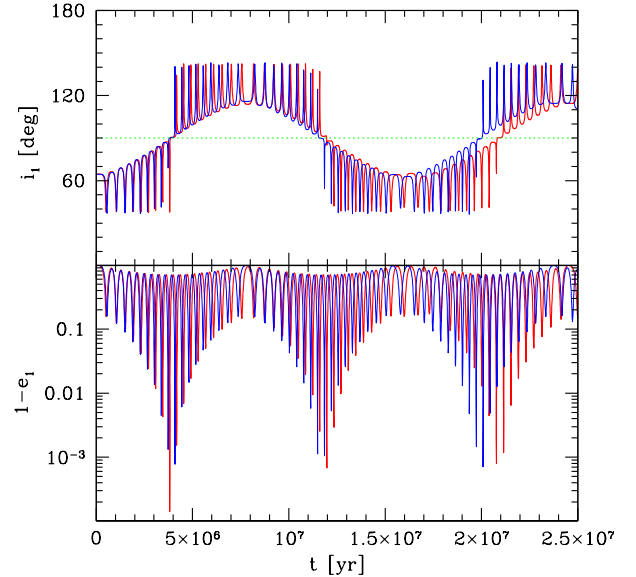}
\caption{Comparison between a direct integration (using a B-S
  integrator) and the octupole-level approximation (see Appendix
  \ref{sec:8}).  The red lines are from the integration of the
  octupole-level perturbation equations, while the blue lines are from
  the direct numerical integration of the three-body system.  Here the
  inner binary contains a star of mass $1\,M_\odot$ and a planet of
  mass $1\,M_{\rm J}$, while the outer object is a brown dwarf of mass
  $40\,M_{\rm J}$.  The inner orbit has $a_1 = 6\,$AU and the outer
  orbit has $a_2 = 100\,$AU. The initial eccentricities are $e_1
  =0.001$ and $e_2 =0.6$ and the initial relative inclination
  $i_\tot=65^\circ$.  The thin horizontal line in the top panel marks
  the $90^\circ$ boundary, separating prograde and retrograde
  orbits. The initial mutual inclination of $65^\circ$ corresponds to
  an inner and outer inclination with respect to the total angular
  momentum (parallel to z) of $64.7^\circ$ and $0.3^\circ$,
  respectively.  Here, the arguments of pericenter of the inner orbit is set to $g_1=45^\circ$ and the outer
  orbit set to zero initially. The SMA of the two orbits (not
  shown) are nearly constant during the direct integration, varying by
  less than $0.02$ percent.  The agreement in both period and
  amplitude of oscillation between the direct integration and the
  octupole-level approximation is quite good.  } \label{fig:compare}
\end{center}
\end{figure}

Given the large, qualitative changes in behavior moving from
quadrupole to octupole order in the Hamiltonian, is it possible that
similar changes in the secular evolution may occur at even higher
orders? The answer to this question probably lies in the elimination
of $G_2$ as an integral of motion at octupole order, leaving only four
integrals of motion: the energy of the system, and the three
components of the total angular momentum.  There are no more integrals
of motion to be eliminated, and thus one might expect no more dramatic
changes in the evolution when moving to even higher orders. It is
possible to see this quantitatively for specific initial conditions
through comparisons with direct $n$-body integrations.  We compared
our octupole equations with direct $n$-body integrations, using the
{\tt Mercury} software package \citep{Mercury}. We used both
Burlisch-Stoer and symplectic integrators \citep{WH91} and found
consistent results between the two. We present the
results of a typical integration compared to the integration of the
octupole-level secular equations in Figure~\ref{fig:compare}.  The
initial conditions (see caption) for this system are those of
\citet[][]{Naoz10}, Figure~1.  We find good agreement between the
direct integration and the secular evolution at octupole order. Both
show a beat-like pattern of eccentricity oscillations, suggesting an
interference between the quadrupole and octupole terms, and both
methods show similar flips of the inner orbit.

\section{Implications and Comparison with Previous Studies}\label{sec:implications}

The \citet{Kozai} and \citet{Lidov} equations of motions are correct
to quadrupole order and for a test particle, but differ from the
correct evolution equations for non-test-particle inner orbits and/or
at octupole order.  In this Section we show how these differences give
rise to qualitatively different evolutionary behaviors than those
assumed in some previous works.

\subsection{Massive Inner Object at the Quadrupole Level}

The danger with working in the wrong limit is apparent if we consider
an inner object that is more massive then the outer object.  While the
TPQ formalism incorrectly assumes that the orbit of the outer body is
fixed in the invariable plane, and therefore the inner body's vertical
angular momentum is constant, the quadrupole-level equations presented
in Appendix \ref{sec:quad_eq_motion} do not.

\begin{figure}
\begin{center}
\includegraphics[width=84mm]{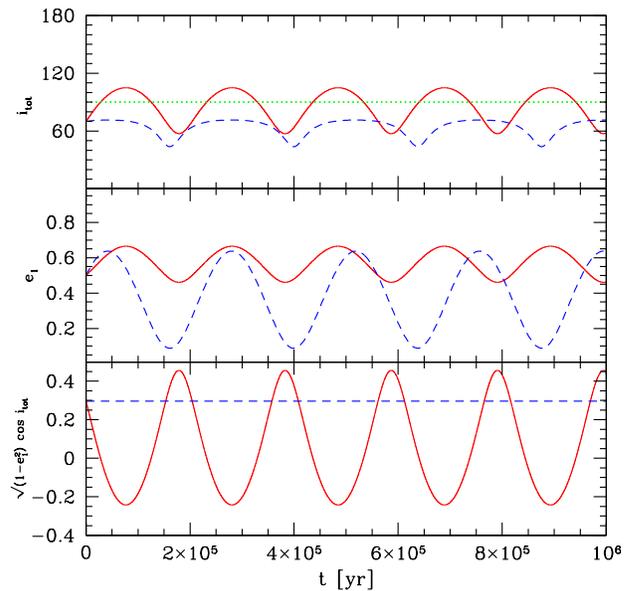}
\caption{Comparison between the standard TPQ formalism (dashed blue
  lines) and our method (solid red lines) for the case of
  PSR~B1620$-$26. Here the inner binary is a millisecond pulsar of
  mass $1.4\,M_\odot$ with a companion of $m_2=0.3\msun$, and the
  outer body has mass $m_3=0.01\msun$. The inner orbit has $a_1 =
  5\,$AU and the outer orbit has $a_2 = 50\,$AU \citep{Ford00Pls}. The
  initial eccentricities are $e_1 =0.5$ and $e_2 =0$ and the initial
  relative inclination $i_\tot=70^\circ$.  The thin horizontal line in
  the top panel marks the $90^\circ$ boundary, separating prograde and
  retrograde orbits. The initial mutual inclination of $70^\circ$
  corresponds to an inner and outer inclination with respect to the
  total angular momentum (parallel to $\hat{\mathbf{z}}$) of
  $6.75^\circ$ and $63.25^\circ$, respectively. The argument of
  pericenter of the inner orbit is initially set $120^\circ$, while
  the outer orbit's is set to zero. We consider, from top to bottom,
  the mutual inclination $i_\tot$, the inner orbit's eccentricity and
  $\sqrt{1-e_1^2} \cos i_\tot$, which the standard formalism assumes
  to be constant (dashed line).  } \label{fig:quad}
\end{center}
\end{figure}

We compare the two formalisms in Figure \ref{fig:quad}.  We consider
the triple system PSR~B1620$-$26 located near the core of the globular
cluster M4.  The inner binary contains a millisecond radio pulsar of
$m_1=1.4 \msun$ and a companion of $m_2=0.3\msun$
\citep{ML88}. Following \citet{Ford00Pls}, we adopt parameters for the
outer perturber of $m_3=0.01\, \msun$ and $e_2=0$. Note that
\citet{Ford00Pls} found $e_2=0.45$, but it is interesting to show that
even for an axisymmetric outer potential the evolution of the system
is qualitatively different then the TPQ approximation  (see the caption for a full description of the initial conditions). 
%
%The inner
%binary separation is $a_1=5$~AU while $a_2=50$~AU. 
%We initialize the
%system with $i_\tot=70^\circ$ and
%$e_1=0.5$ {\bf (see the caption for the full initial conditions)}. 
Note that the actual measured inner binary eccentricity is
$e_1\sim 0.045$, however in order to illustrate the difference we
adopt a higher value ($e_1=0.5$ ).  
%The initial mutual inclination of $70^\circ$
%corresponds to an inner and outer inclination with respect to the
%total angular momentum (parallel to z) of $6.75^\circ$ and
%$63.25^\circ$, respectively.
  For these initial conditions
$\epsilon_M=0.036$, so a careful analysis would require incorporating
the octupole--order terms in the motion; nevertheless, we consider the
evolution of the system to quadrupole order for comparison with the
TPQ formalism.  We have verified, however, that the neglected
octupole--order effects do not qualitatively change the behavior of
the system.  This is because the outer companion mass is low, and
hence the inner orbit does not exhibit large amplitude
oscillations\footnote{Unlike the test particle octupole-level
  approximation \citep{LN,Boaz2}, backreaction of the outer orbit may
  suppress the eccentric Kozai effect. We address this in further
  detail in Teyssandier et al.~(in prep). }.

For the comparison, we do not compare the (constant) $H_1$ from the
TPQ formalism to the (varying) $H_1$ of the correct formalism.
Instead, we compare  the (varying) $H_1$ from the correct formalism (solid red line) with
$G_1 \cos i_\tot$ (dashed blue line), which is the vertical angular momentum that would
be inferred in our formalism \emph{if the outer orbit were
  instantaneously in the invariable plane}, as assumed in the TPQ
formalism.

In Figure \ref{fig:quad}, the mutual inclination oscillates between
$106.7^\circ$ to $57.5^\circ$, and thus crosses $90^\circ$.  These
oscillations are mostly due to the oscillations of the outer orbit's
inclination, while $i_1$ does not change by more than $\sim 1^\circ$
in each cycle. Clearly, the outer orbit does not lie in the fixed
invariable plane.  Figure \ref{fig:quad}, bottom panel, shows
$\sqrt{1-e_1^2}\cos i_\tot$, which, in the TPQ limit, is the vertical
angular momentum of the inner body.

\begin{figure}
\begin{center}
\includegraphics[width=84mm]{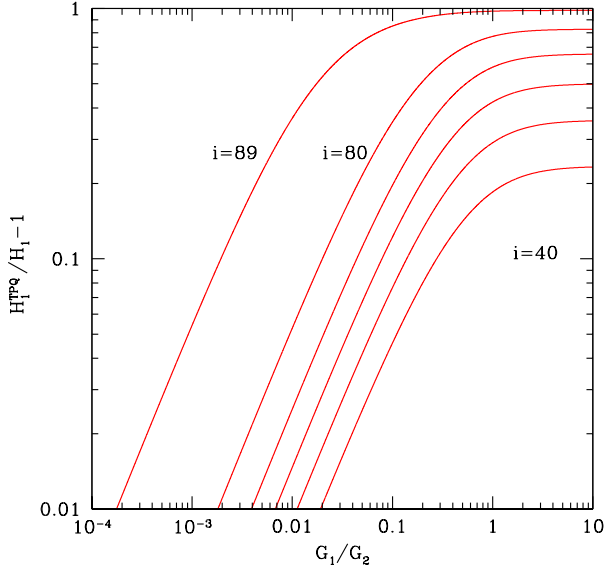}
\caption{The ratio between the correct, changing $z$-component of the
  angular momentum, $H_1$, and the TPQ assumption often used in the
  literature, $H_1^{TPQ}=G_1 \cos i_\mathrm{tot}$.  This ratio was
  calculated analytically for various total angular momentum ratios,
  $G_1/G_2$, and inclinations.  The curves, from bottom to top, have
  $i=40,50,60,70,80$ and $89$ degrees.} \label{fig:H1}
\end{center}
\end{figure}

We can evaluate analytically the error introduced by the application
of the TPQ formalism to this situation.  We compare the vertical
angular momentum ($H_1$) as calculated here to
$H_1^{TPQ}=L_1\sqrt{1-e_1^2} \cos i = {\rm const.}$.  The relative
error between the formalisms is $H_1^{TPQ}/H_1-1$.  In Figure
\ref{fig:H1} we show the ratio between the inner orbit's vertical
angular momentum in the TPQ limit (i.e., $H_1^{TPQ}=G_1 \cos i$) and
equation \eqref{eq:HdotsolF} as a function of the total angular
momentum ratio, $G_1/G_2$, for various inclinations.  Note that this
error can be calculated without evolving the system by using angular
momentum conservation, Eq.~\eqref{eq:cosi}.  The TPQ limit is
only valid when $G_1/G_2 \lsim 10^{-4}$.

\subsection{Octupole--Level Planetary Dynamics}\label{sec:planet}

Recent measurements of the sky-projected angle between the orbits of
several hot Jupiters and the spins of their host stars have shown that
roughly one in four is retrograde \citep{GW07,Tri+10,Albrecht+12}.  If these
planets migrated in from much larger distances through their
interaction with the protoplanetary disk \citep{Lin+86,Mass+03}, their
orbits should have low eccentricities and inclinations\footnote{This
  assumption can be invalid if there are significant magnetic
  interactions between the star and the protoplanetary disk
  \citep{Lai+10} or if there are interaction with another star in a stellar cluster   \citep[e.g.][]{Thies+11,Boley+12} or if there is an episode of planet-planet scattering
  following planet formation \citep{Sourav+08,Nag+08} see also
  \citet{Mer+09}.}.  Disk migration scenarios therefore have
difficulty accounting for the observed retrograde hot Jupiter orbits.
An alternative migration scenario that can account for the retrograde
orbits is the secular interaction between a planet and a binary
stellar companion \citep{Wu+03,Dan,Wu+07,Takeda,Cor+11}.  For an
extremely distant and massive companion ($\epsilon_M  \to \epsilon \ll 1$) the quadrupole
test-particle approximation applies, and $\sqrt{1-e_1^2}\cos i_1$ is
nearly constant (where  the planet is the massless body). Although this forbids orbits that are truly
retrograde (with respect to the total angular momentum of the system),
if the inner orbit begins highly inclined relative to the outer star's
orbit and aligned with the spin of the inner star, then the
star-planet spin-orbit angle can change by more than $90^\circ$ during
the secular evolution of the system, producing apparently retrograde
orbits \citep{Dan,Cor+11}.  Nonetheless, a difficulty with this
``stellar Kozai'' mechanism is that even with the most optimistic
assumptions it can only produce $\lsim 10\%$ of hot Jupiters
\citep{Wu+07}.

\begin{figure}
\begin{center}
\includegraphics[width=84mm]{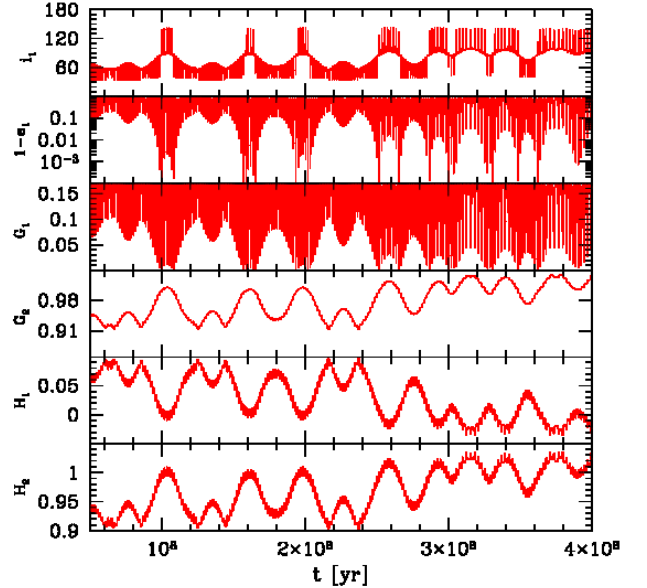}
\caption{Evolution of a planetary system with $m_1=1\,\msun$,
  $m_2=1\,M_J$ and $m_3=2\,M_J$, with $a_1=4$~AU and $a_2=45$~AU. We
  initialize the system at $t=0$ with $e_1=0.01$, $e_2=0.6$,
  $g_1=180^\circ$, $g_2=0^\circ$ and $\itot=67^\circ$.  For these
  initial conditions $i_1=57.92^\circ$ and $i_2=9.08^\circ$. The
  $z$-components of the orbital angular momenta, $H_1$ and $H_2$, are
  shown normalized to the total angular momentum of each orbit.  The
  inner orbit flips repeatedly between prograde ($i_1 < 90^\circ$) and
  retrograde ($i_1 > 90^\circ$). } \label{fig:hypoplanet3}
\end{center}
\end{figure}

\begin{figure}
\begin{center}
\includegraphics[width=84mm]{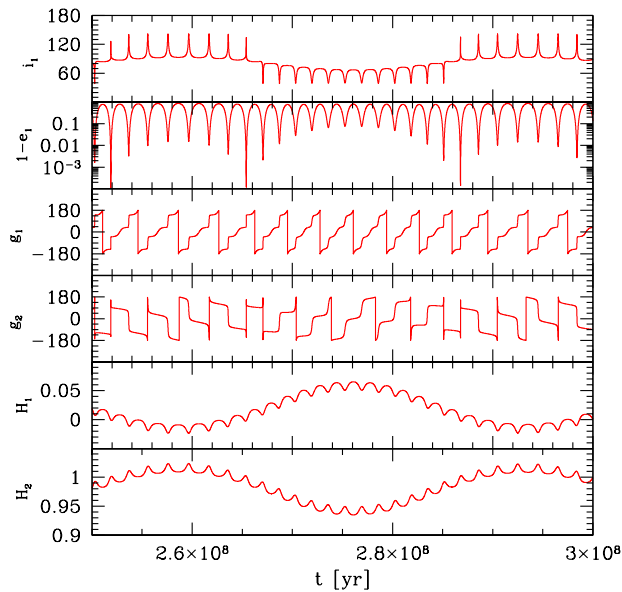}
\caption{Zoom-in on part of the evolution of the point-mass planetary
  system in Figure \ref{fig:hypoplanet3}. In this zoom-in, we can see
  that flips in the inner orbit---$i_1$ crossing $90^\circ$---are
  associated with excursions to very high
  eccentricity.} \label{fig:hypoplanet}
\end{center}
\end{figure}

\citet{Wu+03}, \citet{Wu+07}, \citet{Dan} and \citet{Cor+11} studied
the evolution of a Jupiter-mass planet in stellar binaries in the TPQ
formalism.  For example, the case of HD 80606b
(\citet[Fig. 1]{Wu+03,Dan} and \citet[also Fig. 1]{Cor+11}) was
considered with an outer stellar companion at $1000$~AU.  However, if
the companion is assumed to be eccentric $\epsilon_M$ is not
negligible, and the system is more appropriately described with the
test particle octupole--level approximation \citep[e.g.,][]{LN,Boaz2}.
Furthermore, the statistical distribution for closer stellar binaries
in \citet{Wu+07} and \citet{Dan} is only valid in the approximation
where the outer orbit's eccentricity is zero. In fact, for the systems
considered in those studies $\epsilon_M$ is not negligible and the
octupole--level approximation results in dramatically different
behavior as was shown in  \citet{Naoz+12}. The same dramatic
difference in behavior also exists in the analysis of triple stars
\citep[e.g.,][]{Dan,PF09}, see \S \ref{sec:3S}. 

{A dramatic difference between the octupole and quadrupole--level of approximation is that the  former often generates extremely 
high eccentricities.
In real systems, 
such high eccentricities can be suppressed by 
 tides or GR  \citep[e.g.,][]{Sod84,1998EKH,1998KEM,Borkovits+04}.
Flips can also be prevented because they typically occur shortly after
extreme  eccentricities 
 (see Teyssandier et al. in prep.). 
In our previous studies  that include
tides, planetary perturbers  typically
  allow  flips to happen, while stellar perturbers mostly suppress them \citep{Naoz10,Naoz+12}  But in both cases, tides quantitatively affect the evolution. }

\citet{Naoz10} considered planet-planet secular interactions with tidal interactions as a
possible source of retrograde hot Jupiters.  In this situation
$\epsilon_M$ is not small, requiring computation of the octupole-level
secular dynamics.  In Figures~\ref{fig:hypoplanet3}
and~\ref{fig:hypoplanet} we show the evolution of a representative
configuration   (see the caption for a full description of the initial conditions).
%where $m_1=1$~M$_\odot$, $m_2=1\,M_J$ and $m_3=2\,M_J$,
%with $a_1=4$~AU and $a_2=45$~AU. 
 For this configuration,
$\epsilon_M=0.083$.  Flips of the inner orbit are associated with
evolution to very high eccentricity (see Figures \ref{fig:hypoplanet3} and \ref{fig:hypoplanet}).

\subsection{Octupole--Level  Solar System Dynamics}

\citet{Kozai} studied the dynamical evolution of an asteroid due to
Jupiter's secular perturbations. He assumed that Jupiter's
eccentricity is strictly zero.  However, Jupiter's eccentricity is
$\sim0.05$, and thus studying the evolution of a test particle in the
asteroid belt ($a_1\sim 2-3$~AU) places the evolution in a regime
where the eccentric Kozai-Lidov effect could be significant, with
$\epsilon_M = \epsilon = 0.03$ \citep[][]{LN,Boaz2}.

\begin{figure}
%\begin{center}
\includegraphics[width=84mm]{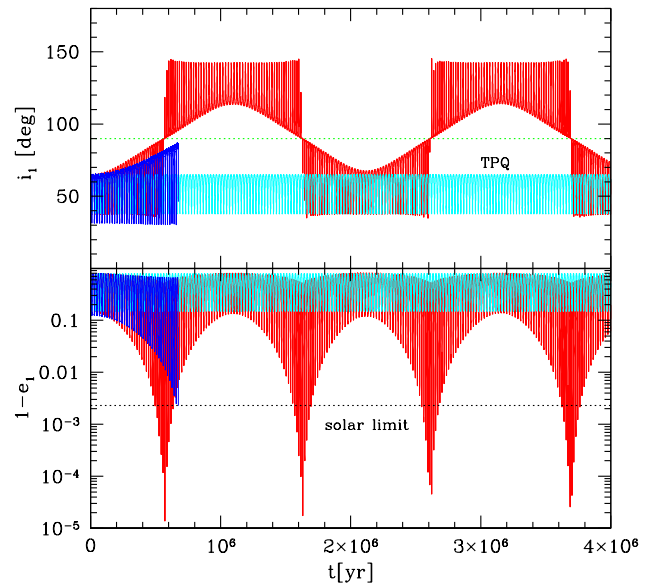}
\caption{Evolution of an asteroid due to Jupiter's secular
  gravitational perturbations (Kozai 1962). We consider $m_1
  =1\,\msun$, $m_2\to0$ and $m_3 =1$~M$_J$, with $a_1 =2$~AU and $a_2
  =5$~AU. We initialize the system at $t=0$ with $e_1 =0.2$, $e_2
  =0.05$, $g_1 =g_2 =0^\circ$ and $i_\tot =65^\circ$. We show the TPQ
  evolution (cyan lines) and the EKL evolution (red lines).  The thin
  horizontal dotted line in the top panel marks the $90^\circ$
  boundary, separating prograde and retrograde orbits.  The inner
  orbit flips periodically between prograde ($i_1 < 90^\circ$) and
  retrograde ($i_1 > 90^\circ$). We also show the result of an
  $N$-body simulation (blue lines). The thin horizontal dotted line in
  the bottom panel marks the eccentricity corresponding to a collision
  with the solar surface, $1-e_1 = R_\odot / a_1$.} \label{fig:Kozai}
\end{figure}

We considered the evolution of asteroid at $2$~AU (assumed to be a
test particle) due to Jupiter at $5$~AU with eccentricity of $e_2 =
0.05$   (see the caption for a full description of the initial conditions). 
% We set $e_1 = 0.2$, $i_\tot = 65^\circ$ and $g_1 = g_2 =
%0^\circ$ initially. 
The asteroid is a test particle and therefore $i_1
\approx i_\tot$. In Figure \ref{fig:Kozai} we compare the evolution of
an asteroid using the TPQ limit \citep[e.g.,][]{Kozai,Thomas+96,KN07}
and the octupole-level evolution discussed here. For this value of
$\epsilon$, the eccentric Kozai-Lidov effect significantly alters the
evolution of the asteroid, even driving it to such high inclination
that the orbit becomes retrograde.  Though we deal only with point
masses in this work, note that the eccentricity is so high that the
inner orbit's pericenter lies well within the sun.

The value of $\epsilon$ here is mainly due to the relative high
$\alpha$ in the problem (an issue raised in the original work on this
problem \citep{Kozai}). The system is very packed which raises
questions with regards to the validity of the hierarchical
approximation.  Even in the EKL formalism, such high eccentricities
occur that the asteriod collides with the sun and the apo-center of
the asteroid approaches about $1$~AU from Jupiter's orbit. To
determine the importance of these effects, we ran an $N$-body
simulation using the {\tt Mercury} software package
\citep{Mercury}. We used both Bulirsch-Stoer and symplectic
integrators \citep{WH91}. The results are depicted at Figure
\ref{fig:Kozai}, which show that the TPQ limit is indeed inadequate
for the system. In addition the octupole--level approximation has some
deviations from the direct $N$-body integration, particularly in the
high eccentricity regime. Note that the evolution of the asteroid in
the direct integration resulted in a collision with the Sun\footnote{As noted in \citet{LN} for very small periapse the integration   becomes extremely costly.}. In
reality, it is likely that a planetary encounter would remove the
asteroid from the solar system before this point.  In contrast to the
EKL mechanism, assuming zero eccentricity for Jupiter results in
consistent results between the secular evolution and the direct
integration \citep{Thomas+96}.

As shown in Figure \ref{fig:Kozai}, taking into account Jupiter's eccentricity
($\sim 0.05$), produces a dramatically different evolutionary
behavior, including retrograde orbits for the
asteroid. \citet{Thomas+96} applied the TPQ formalism to the
asteroid-Jupiter setting (see for example their Figure 2 for $a_1 =
3$~AU). \citet{KN07} developed an analytical solution for the TPQ
limit \citep[see also][]{Kin+91,Kin+99}.

The TPQ formalism has also been applied to the study of the outer
solar system.  \citet{KN07} applied their analytical solution to
Neptune's outer satellite Laomedeia.  This system has $\epsilon\to 0$
and thus the TPQ limit there is justified.  In addition, \citet{PN09}
have studied the evolution of binary minor planets using the TPQ
approximation. In this problem $\epsilon \to 0$ and thus the TPQ
approximation is valid.

\citet[sections 3--4]{Lidov+76} also solved analytically the
quadrapole--level approximation but, unlike \citet{KN07}, they did not
restrict themselves to the TPQ limit, and used the total angular
momentum conservation law in order to calculate the inclinations.
Thus, their formalism is equivalent to ours at quadrupole--order.
Later, \citet{Mazeh+79} also derived evolution equations outside the
TPQ limit (their eqs. A1-A8), allowing for small eccentricities and
inclinations of the outer body.

\subsection{Octupole--Level Perturbations in  Triple Stars}\label{sec:3S}

The evolution of triple stars has been studied by many authors using
the standard (TPQ) formalism
\citep[e.g.,][]{Mazeh+79,1998EKH,1998KEM,Mik+98,Egg+01,Dan,PF09}.  In
some cases the corrected formalism derived here can give rise to
qualitatively different results. We show that some of the previous
studies should be repeated in order to account for the correct
dynamical evolution, and give one example where the eccentric
Kozi-Lidov mechanism dramatically changes the evolution.

\begin{figure}
\begin{center}
\includegraphics[width=84mm]{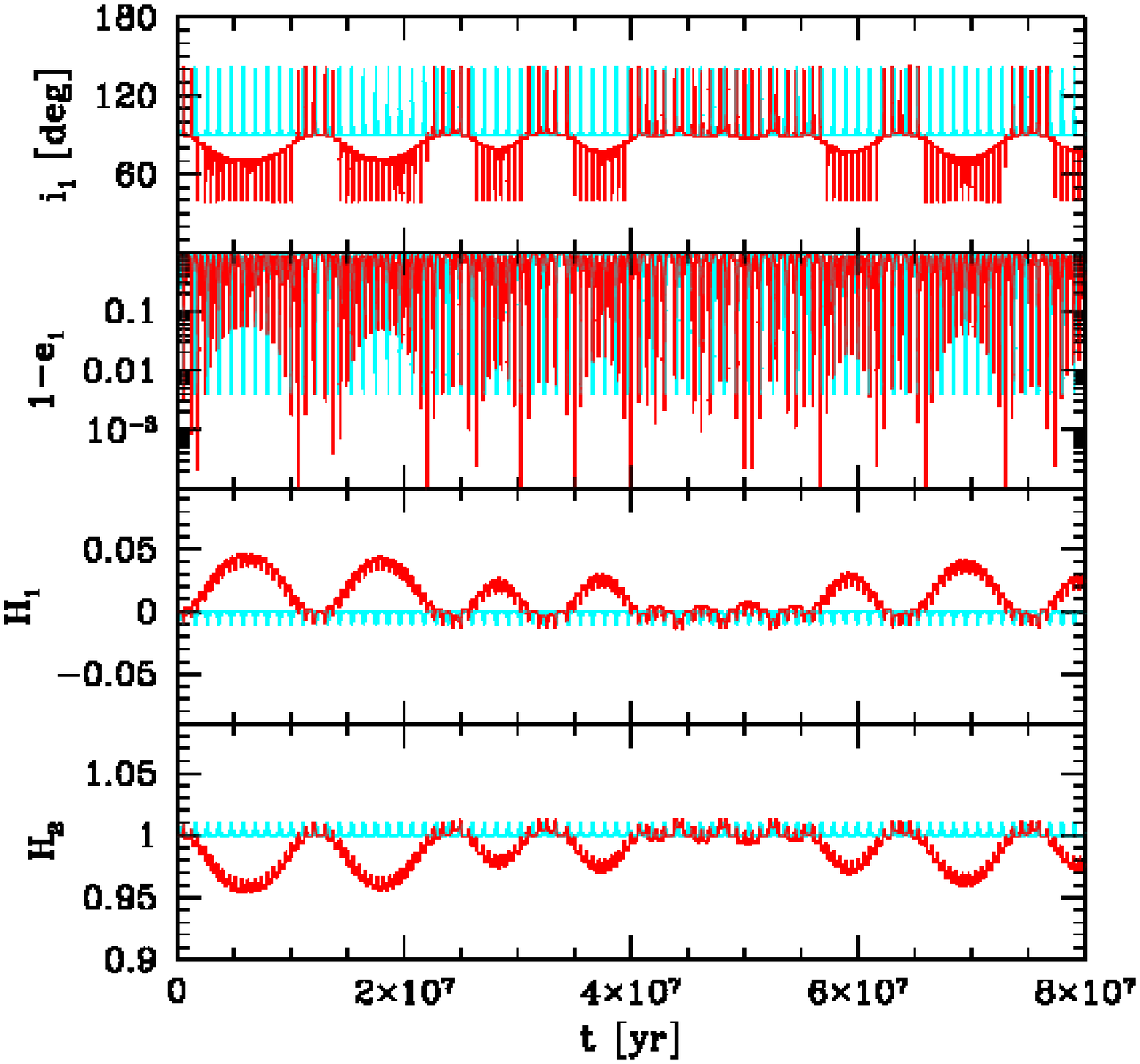}
\caption{An example of dramatically different evolution between the
  quadruple and octupole approximations for a triple-star system. The
  system has $m_1=1\msun$, $m_2=0.25\msun$ and $m_3=0.6\msun$, with
  $a_1=60$~AU and $a_2=800$~AU. We initialize the system with
  $e_1=0.01$, $e_2=0.6$, $g_1=g_2=0^\circ$ and
  $i_\tot=98^\circ$, taken from \citet{Dan}. For these initial
  conditions $i_1=90.02^\circ$ and $i_2=7.98^\circ$.  We show both the
  (correct) quadrupole-level evolution (light-blue lines) and the
  octupole-level evolution (red lines).  $H_1$ and $H_2$, the
  $z$-components of the angular momenta of the orbits, are normalized
  to the total angular momentum.  Note that the octupole-level
  evolution produces periodic transitions from prograde to retrograde
  inner orbits (relative to the total angular momentum), while at the
  quadrupole-level the inner orbit remains
  prograde.  See Figure \ref{fig:hypostar_B} for comparison with direct numerical integration of the three-body system.}\label{fig:hypostar}
\end{center}
\end{figure}

\begin{figure}
\begin{center}
\includegraphics[width=84mm]{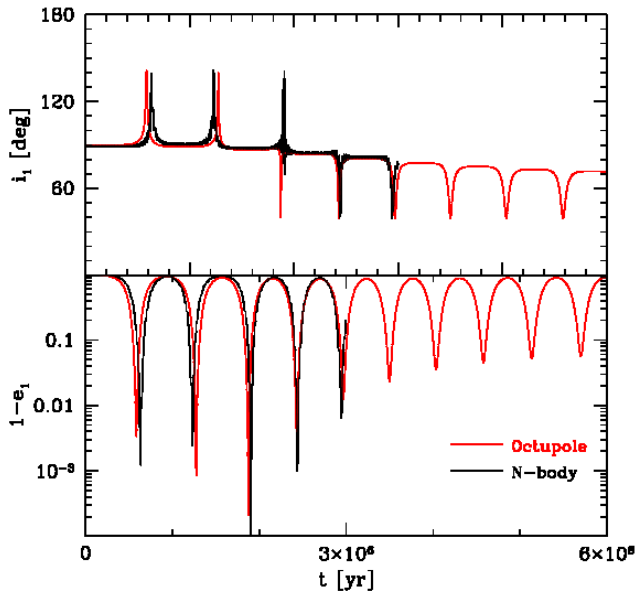}
\caption{ The evolution for the first $6$~Myr, of Figure \ref{fig:hypostar} where we show a comparison between direct 3--body integration (using a B-S
integrator),  and the octupole--level of approximation. The   red lines are from the integration of the octupole-level
perturbation equations, while the black lines are from the direct
numerical integration of the three-body system.}\label{fig:hypostar_B}
\end{center}
\end{figure}

\citet{Dan} studied the distribution of triple star properties using
Monte Carlo simulations.  We choose a particular system from their
triple-star suite of simulations to illustrate how the dynamics
including the octupole order can be qualitatively different from what
would be seen at quadrupole order  (see the caption for a full description of the initial conditions).  
%We adopt the following parameters: $m_1=1 \msun$,
%$m_2=0.25\msun$ and $m_3=0.6\msun$, $a_1=60\,$AU and $a_2=800\,$AU.  We
%initialize the system at $t=0$ with $e_1=0.01$, $e_2=0.6$,
%$g_1=g_2=0^\circ$ and $\itot=98^\circ$, corresponding to
%$i_1=90.02^\circ$, $i_2=7.98^\circ$ and 
For this system $\epsilon_M = 0.042$ (and $\epsilon=0.0703$).  The
evolution of the system is shown in Figure~\ref{fig:hypostar}. At
octupole order, the inclination of the inner orbit oscillates between
about $40^\circ$ and $140^\circ$, often becoming retrograde (relative
to the total angular momentum), while the quadrupole--order behavior is
very different and the inner orbit remains always prograde.  The
octupole--order treatment also gives rise to much higher eccentricities
\citep{KM99,Ford00}.  In Figure~\ref{fig:hypostar_B}  we compare the octupole--level  evolution (of the same system) with direct 3--body integration. 

The evolution shown in Figure~\ref{fig:hypostar}
is for point-mass stars; in reality, these high-eccentricity
excursions would actually drive the inner binary to its Roche limit,
leading to mass transfer. For these high eccentricities tides will play an important role and thus in reality flips in similar systems may be suppressed.  Similarly the high eccentricities often
excited through the eccentric Kozai mechanism can also lead to compact
object binary merger.

The possibility of forming blue stragglers through secular
interactions in triple star systems has been suggested by \citet{PF09}
and \citet{aaron+11}.  As shown in \citet{KM99,Ford00} and in the
example above the minimum pericenter distance of the inner binary can
differ significantly between the TPQ and EKL formalisms.  This
suggests that using the correct EKL formalism could significantly
increase the computed likelihood of such a formation mechanism for
blue stragglers.

\begin{figure}
\begin{center}
\includegraphics[width=84mm]{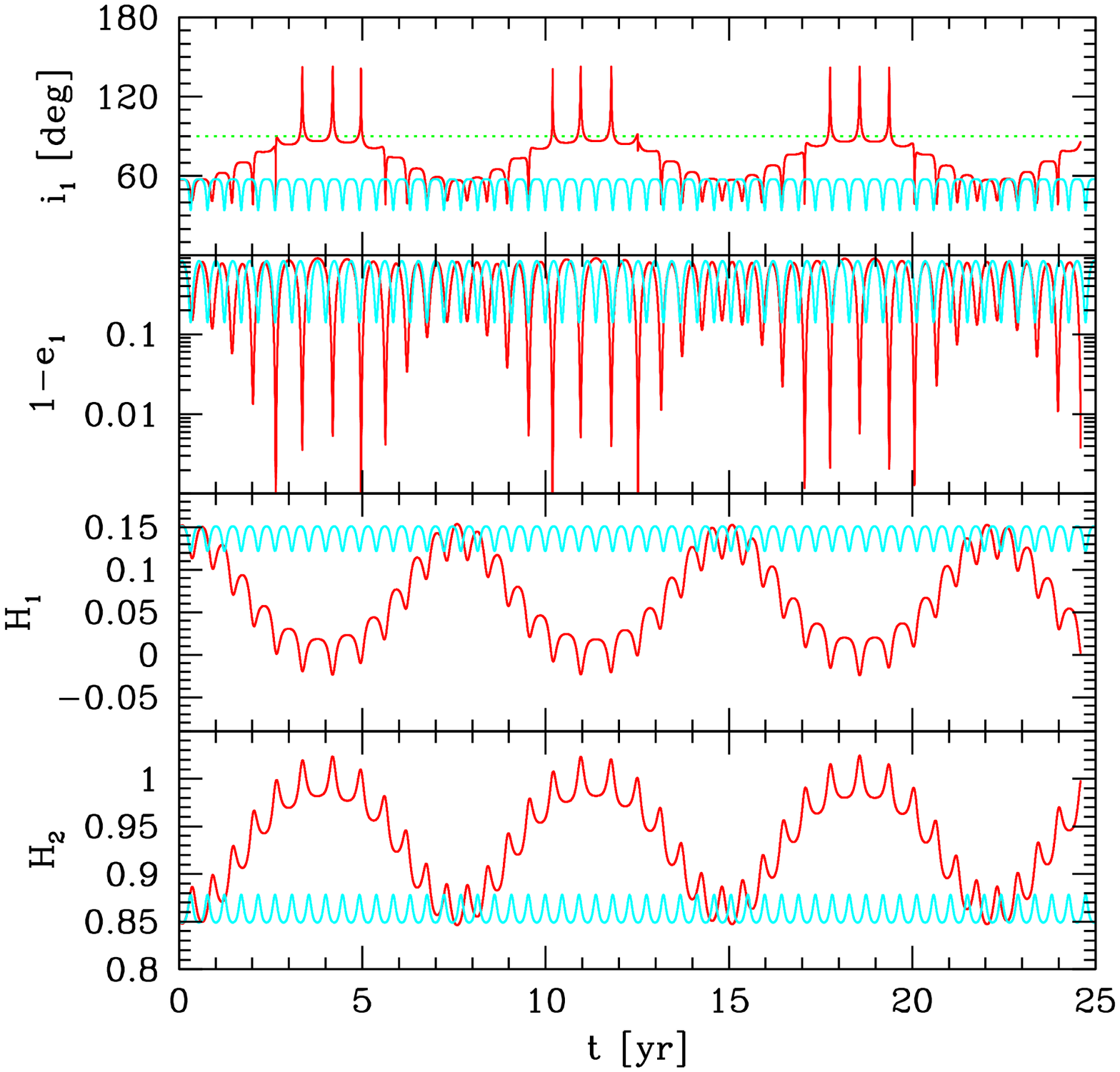}
\caption{An example of dramatically different evolution between the
  quadruple and octupole approximations for a triple star system
  representing the best-fit parameters from the \citet{Mik+98}
  analysis of CH Cygni. The system has $m_1=3.51\msun$, $m_2=0.5\msun$
  and $m_3=0.909\msun$, with $a_1=0.05$~AU and $a_2=0.21$~AU. We
  initialize the system with $e_1=0.32$, $e_2=0.6$, $g_1=145^\circ$,
  $g_2=0^\circ$ and $i_\tot=72^\circ$.  For these initial conditions
  $i_1=57.02^\circ$ and $i_2=14.98^\circ$.  We show both the (non-TPQ)
  quadrupole-level evolution (light-blue lines) and the octupole-level
  evolution (red lines).  $H_1$ and $H_2$, the z-components of the
  angular momenta of the orbits, are normalized to the total angular
  momentum.  Note that the octupole-level evolution produces periodic
  transitions from prograde to retrograde inner orbits (relative to
  the total angular momentum), while at the quadrupole-level the inner
  orbit remains prograde.  To avoid clutter in the figure we have
  omitted the TPQ result.  In the TPQ formalism, the evolution of the
  inclination and eccentricity are similar to the general
  quadrupole-level approximation, but $H_{1,2}$ are constant.
} \label{fig:Cygni}
\end{center}
\end{figure}

For many years CH Cygni was considered to be an interesting triple
candidate because it exhibits two clear distinguishable periods
\citep[e.g.][]{Don+95,Skopal,Mik+98,Hin+93}. % A short period of $\sim
                                             % 2$~days and a longer
                                             % one of $\sim 15$~days.
However, a triple system model based on the TPQ Kozai mechanism
\citep{Mik+98} did not reproduce the observed masses of the system
\citep{Hin+93,Hin+09}.  Applying the corrected formalism in this paper
to the system parameters derived in \citet{Mik+98} gives a very
different evolution than in the TPQ formalism%
\footnote{\citet{Mik+98} also found somewhat different set of
  parameters when producing a fit for data set with less weight for
  the data of 1983 due to large noise in the active phase of the
  system.}. %
Therefore, it seems likely that an analysis based on the formalism
discussed in this paper would give a significantly different fit.  In
Figure \ref{fig:Cygni} we illustrate the differences between the TPQ,
correct quadrupole, and octupole evolution of the system. The best-fit
parameters of the system are taken from \citet{Mik+98} where $\epsilon_M
= 0.14$  (see the caption for a full description of the initial conditions, where we allowed for a freedom in our choice of $e_2,g_1,g_2$ and
$\itot$ since the best fit was found using the TPQ limit, at which
$e_2$ is fixed).
%
%are as follows: $m_1=3.51
%\msun$, $m_2=0.5\msun$ and $m_3=0.909\msun$, $a_1=0.05\,$AU and
%$a_2=0.21\,$AU.  We initialize the system at $t=0$ with $e_1=0.32$,
%$e_2=0.6$, $g_1=145^\circ$, $g_2=0^\circ$ and $\itot=72^\circ$,
%corresponding to $i_1=57.01^\circ$, $i_2=14.98^\circ$ and $\epsilon_M
%= 0.14$.  
%We allowed for a freedom in our choice of $e_2,g_1,g_2$ and
%$\itot$ since the best fit was found using the TPQ limit, at which
%$e_2$ is fixed.  
Note that the choice of the inner eccentricity does
not strongly influence the evolution while the choice of the outer
orbit's eccentricity does. Most importantly, the rather large $\epsilon_M$ for this system implies that the system is not stable, i.e., the averaging over the orbits is not justified. From direct integration we found that the system undergoes strong encounters and the inner binary collides in this example.

\begin{figure}
\begin{center}
\includegraphics[width=84mm]{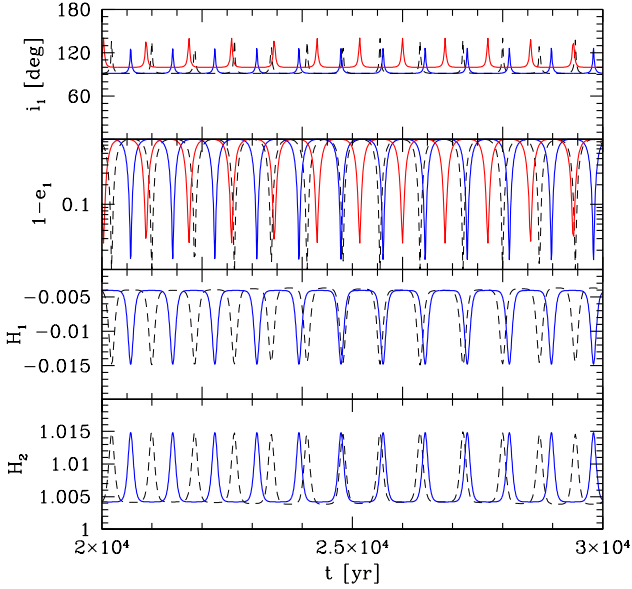}
\caption{The time evolution of  an Algol--like system \citep{1998EKH}, with
  $(m_1,m_2,m_3)=(2.5,2,1.7)$~M$_\odot$. The inner orbit has $a_1 =
  0.095\,$AU and the outer orbit has $a_2 = 2.777\,$AU.  The initial
  eccentricities are $e_1 =0.01$ and $e_2 =0.23$ and the initial
  relative inclination $i_\tot=100^\circ$.  The $z$-components of the
  inner and outer orbital angular momentum, $H_1$ and $H_2$ are
  normalized to the total angular momentum.  The initial mutual
  inclination of $100^\circ$ corresponds to inner- and outer-orbit
  inclinations of $91.6^\circ$ and $8.4^\circ$, respectively.  We
  consider the (correct) quadrupole-level evolution (blue lines),
  octupole-level evolution (dashed lines) and also the standard
  (incorrect) TPQ evolution. In the latter we have assumed, as in
  previous papers, that $i_\tot=i_1$, which results in the discrepancy
  between the inclination values.  See also Figure \ref{fig:AlgolNew}
  for the evolution of the  Algol--like system using the updated masses and
  orbital parameter, following \citet{Baron+12}.} \label{fig:Algol}
\end{center}
\end{figure}

 It is also interesting to investigate a system for which the eccentric
Kozai mechanism is suppressed due to comparable masses for the inner
orbit, and low eccentricity of the outer orbit (i.e., $\epsilon_M<<1$).  \citet{1998KEM} and \citet{Egg+01} studied the Algol triple
system \citep{Algol} using the TPQ equations.  The TPQ equations were
also used in the paper that introduced the influential KCTF mechanism
\citep{Mazeh+79,1998EKH}. Note that tides dominate the evolution of the Algol system today \citep[e.g.,~][]{sod06}.  Figure~\ref{fig:Algol} compares the
evolution computed in the (incorrect) TPQ formalism, the correct
quadrupole formalism, and the octupole-level EKL formalism applied to
 an Algol--like system.  The correct quadrupole formalism
decreases the minimum value of $1-e_1$ by almost a factor of 2
relative to the TPQ formalism.  The reduced pericenter distance would
strongly increase the effects of tidal friction (not included here),
which may lead to rapid circularization of the inner orbit.  The
octupole-level computation decreases the minimum pericenter distance
by a further 40\%.

\begin{figure}
\begin{center}
\includegraphics[width=84mm]{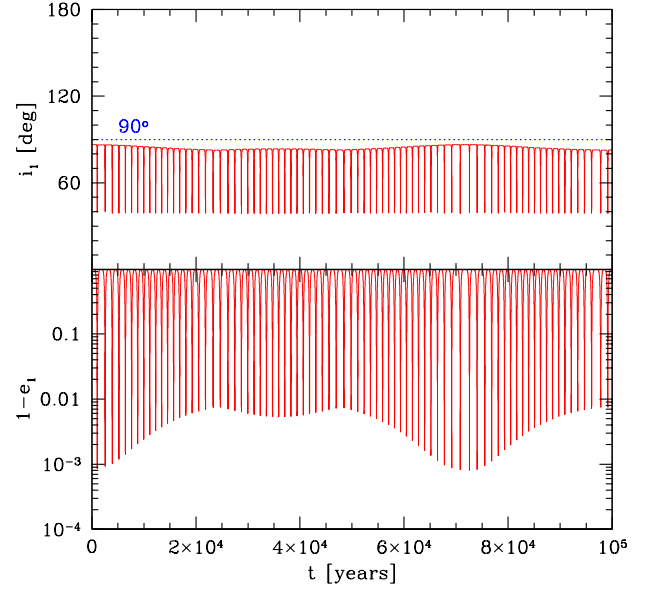}
\caption{ The time evolution of  Algol--like  system using the orbital
  parameters taken from \citet{Baron+12}, with
  $(m_1,m_2,m_3)=(3.17,0.7,1.7)$~M$_\odot$. The inner orbit has $a_1 =
  0.062\,$AU and the outer orbit has $a_2 = 2.68\,$AU.  The initial
  eccentricities are $e_1 =0.001$ and $e_2 =0.23$ and the initial
  relative inclination $i_\tot=90^\circ$.  The initial mutual
  inclination of $90^\circ$ corresponds to inner- and outer-orbit
  inclinations of $86.4^\circ$ and $3.6^\circ$, respectively.  We
  consider only the octupole-level evolution. Compare this to the
  evolution  in Figure \ref{fig:Algol}.} \label{fig:AlgolNew}
\end{center}
\end{figure}

Note that the masses and orbital parameters used in \citet{1998KEM}
and \citet{Egg+01} are out of date.  New observations
\citep[e.g.,][]{Baron+12} find the secondary mass to be smaller then
the primary and the mutual inclination to be closer to $90^\circ$. In
Figure \ref{fig:AlgolNew} we show the octupole--level evolution of the
system considering the new parameters. In the absence of any
additional physical mechanism, such as general relativity, tides, mass
transfer, etc., the EKL mechanism could play a very important role in
the dynamical evolution of the system.

 We note that the inner binary in the Algol system is dominated by tidal effects  \citep{Sod75,1998KEM,Egg+01} and figures \ref{fig:Algol} and \ref{fig:AlgolNew}  do not represent the system today  but an Algol--like analogy. We use the Algol parameters here only to show hypothetical outcomes of the correct dynamical evolution. It would be interesting to study  stellar evolution including  tides in the context of the EKL mechanism, for a system such as Algol. 

\subsection{The Danger of the Quadrupole--Level of Approximation }

The octupole-level Hamiltonian and equations of motion were previously
derived by \citet{Har68,Har69,Sid83,Maechal90,KM99,Ford00,Bla+02} and
\citet{Lee03}. Most of the equations of motion can be derived
correctly when applying the elimination of the nodes---only the
$\dot{H}_1$ and $\dot{H}_2$ equations are affected.  These authors
calculated the time evolution of the inclinations (i.e.\ $H_1$ and
$H_2$) from the {\it total} (conserved) angular momentum, and thus
avoided the problem that arises when eliminating the nodes from the
Hamiltonian.  In appendix \ref{sec:8} we show the complete set of
equations of motion for the octupole-level approximation, derived from
a correct Hamiltonian, including the nodal terms.

As displayed here the octupole-level approximation gives rise to a
qualitatively different evolutionary behavior for cases where
$\epsilon_M$ [see eq.~(\ref{eq:epsiM})] is not negligible. We note
that many previous studies applied the quadrupole-level approximation,
which may lead to significantly different results
\citep[e.g.,][]{Mazeh+79,Quinn+90,Bailey+92,Inn+97,1998EKH,Mik+98,Egg+01,3book,Dan,Wu+07,Zdz+07,Takeda,PF09}.
Neglecting the octupole-level approximation can cause changes in the
dynamics varying from a few percent to completely different
qualitative
behavior.

Some other derivations of octupole--order equations of motion dealt
with the secular dynamics in a general way, without using Hamiltonian
perturbation theory or elimination of the nodes
\citep{Far+10,Las+10,Mar10,Boaz}.  In these works there were no
references to the discrepancy between these derivations and the
previous studies. Also, note that the results of \citet{hol+97} are
based on a direct N-body integration, and thus are not subject to the
errors mentioned above.

\section{Conclusions}\label{Sec:con}

We have shown that the ``standard'' TPQ Kozai formalism
\citep{Kozai,Lidov} has been applied in inappropriate situations.  A
common error in the implementation of the relevant Hamiltonian
mechanics (premature elimination of the nodes) leads to the
(incorrect) conclusion that the conservation of the $z$-component of each
orbit's angular momentum from the TPQ dynamics generalizes beyond the
TPQ approximation.  Correcting the formalism we find that the
$z$-components of  \emph{both} the inner and outer orbits' angular
momenta in general change with time at both the quadrupole and
octupole level.  The conservation of the inner orbit's $z$-component
of the angular momentum (the famous $\sqrt{1-e_1^2}\cos i={\rm
  constant}$) only holds in the quadrupole-level \emph{test particle}
approximation.  We have explained in details the source of the error
in previous derivations (Appendix \ref{prob}).

We have re-derived the secular evolution equations for triple systems
using Hamiltonian perturbation theory to the octupole-level of
approximation (Section \ref{Sec:Ham} and Appendix \ref{Sec:form},
\ref{sec:H8} and Appendix \ref{sec:8}). We have also shown that one
can use the simplified Hamiltonian found in the literature
\citep[e.g.,][]{Ford00} as long as the equations of motion for the
inclinations are calculated from the total angular momentum.

The correction shown here has important implications to the evolution
of triple systems. We discussed a few interesting implications in
Section \ref{sec:implications}. We showed that already at the
quadrupole-level approximation the explicit assumption that the
vertical angular momentum is constant can lead to erroneous results,
see for example Figure \ref{fig:quad}. In this Figure we showed that
far from the test particle limit in the quadrupole-level one can
already find a significant difference in the evolutionary
behavior. The correct results agree with the test particle limit only when  $G_1/G_2< 10^{-4}$ (see Figure \ref{fig:H1}).  We
show in Appendix \ref{sec:maxmin} that at the quadrupole level of
approximation, the inner eccentricity and the mutual inclination have
a well defined maximum and minimum irrespective of the mass of the
inner bodies. In the test particle limit these values converge to the
well-known critical inclinations ($39.2^\circ \leq i_0 \leq 140.8^\circ 
$) for large oscillatory amplitudes.

{The most notable outcome of the results presented here happens in the
octupole-level of approximation (which we call the EKL formalism),
when the inner orbit flips from prograde to retrograde with respect to
the total angular momentum.  Just before the flip the inner orbit has an excursion of extremely high eccentricity. In the presence of tidal forces (not included in this study) the outcome of a system can be  different than the one assumed while using the TPQ formalism.
 \citet{KM99}, \citet{Ford00}, \citet{Bla+02}, \citet{Lee03} and
\citet{Las+10}  present the 
   correct octupole equations of motion.  Had these authors integrated their
   equations for systems
   such as those presented in this paper, they could already have discovered the possibility of flipping the inner orbit.
}

%
%had the correct equations of motion, and could, in
%principle, have observed this phenomena.  However, it seems that the
%assumption of a constant vertical angular momentum was built into the
%community understanding of Kozai mechanism, causing the flipping
%effect of the EKL mechanism to be overlooked.
%   
% In \citet{Naoz10} we suggested that this effect may play an important
% role in the formation mechanism of retrograde Hot Jupiters. There we
% showed the importance of this effect in a verity of planetary and
% stellar triple systems. For some examples see Figures
% \ref{fig:hypoplanet3}--\ref{fig:hypostar}, and \citet{Naoz10} where we
% specifically discussed the evolution of two planet systems, triple
% stars and asteroids due to gravitational perturbations from Jupiter.
% We also compared our derivation with direct N-body integration and
% illustrated the same qualitative evolution.  We also emphasized the
% importance of higher-orders approximations, where $\epsilon_M$ is
% significant.

\section*{Acknowledgments}
We thank Boaz Katz, Rosemary Mardling and  Eugene Chiang for useful discussions. We thank Staffan S$\ddot{\rm o}$derhjelm, our referee for very useful comments the improved the manuscript in great deal. We also thank Keren Sharon
and Paul Kiel for comments on the manuscript.   S.N.  supported by NASA through a Einstein Postdoctoral Fellowship awarded by the Chandra X-ray Center, which is op- erated by the Smithsonian Astrophysical Observatory for NASA under contract PF2-130096. Y.L. acknowledges support from NSF grant AST-1109776. Simulations for this project were performed on the HPC
cluster {\it{fugu}} funded by an NSF MRI award.

\bibliographystyle{natbib}

\bibliography{Kozai}
%\begin{thebibliography}{}

%\end{thebibliography}
\appendix
%%%%%%%%%%%%%%%%%%%%%%%%%%%%%%%
%%%%%%%%%%%%%%%%%%%%%%%%%%%%%%%
\section{  The Quadrupole level of Approximation }\label{Sec:form}
%{\sn {I added a sub appendix here which contains the equations of motion of the quadrupole level}}

We develop the complete quadrupole-level secular approximation in this
section.  As mentioned, the main difference between the derivation
shown here and those of previous studies lies in the ``elimination of
nodes'' \citep[e.g.,][]{Kozai,JM66}, which relates to the transition
the \emph{invariable plane} \citep[e.g.,][]{MD00} coordinate system,
where the total angular momentum lies along the $z$-axis.

\subsection{ Transformation to the Invariable Plane }

We choose to work in a coordinate system where the total initial
angular momentum of the system lies along the $z$ axis (see Figure
\ref{fig:angular}),; the $x$-$y$ plane in this coordinate system is
known as the \emph{invariable plane} \citep[e.g.,][]{MD00}, and
therefore we call this coordinate system the \emph{invariable
  coordinate system}.  We begin by expressing the vectors ${\bf r}_1$
and ${\bf r}_2$ each in a coordinate system where the periapse of the
orbit is aligned with the x-axis and the orbit lies in the x-y plane,
called the ``orbital coordinate system,'' and then rotating each
vector to the invariable coordinate system.  The rotation that takes
the position vector in the orbital coordinate system to the position
in the invariable coordinate system is given by \citep[see][chapter
2.8, and Figure 2.14 for more details]{MD00}
\begin{equation}
{\bf r}_{1,\textnormal{inv}}=R_z(h_1)R_x(i_1)R_z(g_1){\bf r}_{1,\textnormal{orb}} \ ,
\end{equation}
where the subscript ``inv'' and ``orb'' refer to the invariable and
orbital coordinate systems, respectively.  The rotation matrices $R_z$
and $R_x$ as a function of rotation angle, $\theta$, are
\begin{equation}
R_z(\theta) = \left(
\begin{array}{ccc}
\cos\theta & -\sin\theta & 0 \\
\sin\theta & \cos\theta & 0 \\
0 & 0 & 1
\end{array}\right)
\end{equation}
and
\begin{equation}
R_x(\theta) = \left(
\begin{array}{ccc}
1 & 0 & 0 \\
0 &\cos\theta & -\sin\theta \\
0 &\sin\theta & \cos\theta
\end{array}\right) \ .
\end{equation}
Thus, the angle between ${\bf r}_1$ and ${\bf r}_2$ is given by:
\begin{equation}\label{eq:cosphi}
  \cos\Phi = \hat{{\bf r}}_{2,\textnormal{orb}}^T R^{-1}_z(g_2) R^{-1}_x(i_2) R_z^{-1}(h_2) R_z(h_1)
  R_x(i_1) R_z(g_1) \hat{{\bf r}}_{1,\textnormal{orb}},
\end{equation}
where $\hat{{\bf r}}_{1,2,\textnormal{orb}}$ are unit vectors that
point along ${\bf r}_{1,2,\textnormal{orb}}$.  In the orbital
coordinate system, we have
\begin{equation}
  \label{eq:orbit-plane-rhat}
  \hat{{\bf r}}_{1,2,\textnormal{orb}} = \begin{pmatrix}
    \cos\left( f_{1,2} \right) \\
    \sin\left( f_{1,2} \right) \\
    0
\end{pmatrix}\, ,
\end{equation}
where $f_{1}$ ($f_2$) is the true anomaly for the inner (outer) orbit.
Note that $R_z^{-1}(h_2) R_z(h_1)=R_z(h_1-h_2) \equiv R_z(\Delta h)$,
so the Hamiltonian will depend on the difference in the longitudes of
the ascending nodes; in a similar manner, the Hamiltonian depends on
$f_1$ and $f_2$ only through expressions of the form $f_1 + g_1$ and
$f_2 + g_2$.  Replacing $\cos\Phi$ in the Hamiltonian,
eq.~(\ref{eq:Ham4}), we can now integrate over the the mean anomaly
angles using the Kepler relations between the mean and true anomalies:
\begin{equation}
dl_i=\frac{1}{\sqrt{1-e_i^2}}\left(\frac{r_i}{a_i}\right)^2 df_i \ ,
\end{equation}
where for the outer orbit one should simply replace the subscript ``1"
with ``2".
 
\subsection{
%\x Eliminating Short Term Period Canonical Transformation }
   Transformation to Eliminate Mean Motions }\label{sec:short} 

Because we are interested in the long-term dynamics of the triple
system, we now describe the transformation that eliminates the
short-period terms in the Hamiltonian that depend of $l_1$ and $l_2$.
The technique we will use is known as the Von Zeipel transformation
\citep[for more details, see][]{bro59}.

Write the triple-system Hamiltonian in eq.~\eqref{eq:Ham4} as 
\begin{equation}
  \Ham = \Ham_1^K + \Ham_2^K + \Ham_2,
\end{equation}
where $\Ham_1^K$ and $\Ham_2^K$ are the Kepler Hamiltonians that
describe the inner and outer elliptical orbits in the triple system
and $\Ham_2$ describes the quadrupole interaction between the orbits.
Note that $\Ham_2$ is $\mathcal{O}\left( \alpha^2 \right)$, and is the
only term in $\Ham$ that depends on $l_1$ or $l_2$.  We seek a
canonical transformation that can eliminate the $l_1$ and $l_2$ terms
from $\Ham_2$.  Such a transformation must be close to the identity,
since $\Ham_2 \ll \Ham$; let the generating function be
\begin{eqnarray}
  S(L^*_j, G^*_j, H^*_j, l_j, g_j, h_j) &= &\sum_{j = 1}^2 \left[ L^*_j
    l_j + G^*_j g_j + H^*_j h_j \right] \\
    &+& \alpha^2 S_2(L^*_j, G^*_j, H^*_j, l_j, g_j, h_j) \ , \nonumber
\end{eqnarray}
where we indicate the new momenta with a superscript asterix, and
$S_2$ is the non-identity piece of the transformation that we will use
to eliminate $\Ham_2$.  The relationship between the new and old
canonical variables is
\begin{equation}
  \label{eq:momentum-relation}
  p_i = \frac{\partial S}{\partial q_i} = p^*_i + \alpha^2
  \frac{\partial S_2}{\partial q_i}
\end{equation}
and
\begin{equation}
  \label{eq:coordinate-relation}
  q_i^* = \frac{\partial S}{\partial p^*_i} = q_i + \alpha^2
  \frac{\partial S_2}{\partial p^*_i},
\end{equation}
where the momenta $p_i \in \left\{L_i, G_i, H_i \right\}$, and the
coordinates $q_i \in \left\{l_i, g_i, h_i\right\}$.  Because our
generating function is time-independent, the new and old Hamiltonians
agree when evaluated at the corresponding points in phase space:
\begin{equation}
  \Ham(q_i, p_i) = \Ham^*(q_i^*, p_i^*)
\end{equation}
when the phase space coordinates satisfy
equations~\eqref{eq:momentum-relation} and
\eqref{eq:coordinate-relation}.  Inserting these relations into the
un-transformed Hamiltonian, and expanding to lowest order in
$\alpha^2$, we have
\begin{equation}
  \Ham(q_i^*, p_i^*) + \alpha^2 \frac{\partial \Ham}{\partial p_i}
  \frac{\partial S_2}{\partial q_i} - \alpha^2 \frac{\partial
    \Ham}{\partial q_i} \frac{\partial S_2}{\partial p_i^*} =
  \Ham^*\left( q_i^*, p_i^* \right).
\end{equation}
Equating terms order-by-order in $\alpha$ gives
\begin{equation}
  \Ham^K_1(q_i^*, p_i^*) = \Ham^{*K}_1(q_i^*, p_i^*),
\end{equation}
\begin{equation}
  \Ham^K_2(q_i^*, p_i^*) = \Ham^{*K}_2(q_i^*, p_i^*),
\end{equation}
and
\begin{equation}
  \Ham_2\left( q_i^*, p_i^* \right) + \alpha^2 \sum_{i=1}^2 \frac{\partial \Ham}{\partial p_i}
  \frac{\partial S_2}{\partial q_i} - \alpha^2 \sum_{i=1}^2 \frac{\partial
    \Ham}{\partial q_i} \frac{\partial S_2}{\partial p_i^*} =
  \Ham^*_2\left( q_i^*, p_i^* \right).
\end{equation}
Since the last two terms on the left-hand side of this latter equation
are already $\mathcal{O}\left( \alpha^2 \right)$, only the $\Ham_1^K$
and $\Ham_2^K$ parts of $\Ham$ contribute.  These Kepler Hamiltonians
only depend on $L_1$ and $L_2$, so there are only two non-zero
partials of $\Ham$ at order $\alpha^2$:
\begin{equation}
  \Ham_2\left( q_i^*, p_i^* \right) + \alpha^2 \frac{\partial \Ham_1^K}{\partial L_1}
  \frac{\partial S_2}{\partial l_1} + \alpha^2 \frac{\partial \Ham_2^K}{\partial L_2}
  \frac{\partial S_2}{\partial l_2} =
  \Ham^*_2\left( q_i^*, p_i^* \right).
\end{equation}
We must use the terms that depend on $S_2$ to cancel any terms in
$H_2$ that depend on $l_1^*$ and $l_2^*$.  Note that $\Ham_2$ is
periodic in $l_1^*$ and $l_2^*$ with period $2\pi$ (see
equations~\eqref{eq:cosphi} and \eqref{eq:orbit-plane-rhat}), so we
can write
\begin{equation}
  \Ham_2\left( q_i^*, p_i^* \right) = \alpha^2 h_0 + \alpha^2
  \sum_{k_1, k_2
    = 1}^\infty h_{k_1 k_2}
  e^{-i k_1 l_1^* - i k_2 l_2^*},
\end{equation}
with 
\begin{equation}
  h_{k_1 k_2} = \frac{1}{4\pi^2 \alpha^2} \int_{0}^{2\pi} d l_1^* dl_2^*\, \Ham_2\left( q_i^*,
    p_i^* \right) e^{i k_1 l_1^*+i k_2 l_2^*}.
\end{equation}
Now let $\partial \Ham^K_1 / \partial L_1 \equiv \omega_1(L_1)$, and
$\partial \Ham^K_2 / \partial L_2 \equiv \omega_2(L_2)$.  Suppose that
$S_2$ is periodic in $l_1$ and $l_2$ (which are equivalent, at lowest
order, to $l_1^*$ and $l_2^*$).  Then
\begin{multline}
  \alpha^2 h_0 + \alpha^2 \sum_{k_1,k_2 = 1}^\infty h_{k_1 k_2}
  e^{-i k_1 l_1^* - i k_2 l_2^*} + \alpha^2 \omega_1
  \sum_{k_1,k_2=1}^{\infty} -i k_1 s_{k_1 k_2}
  e^{-i k_1 l_1 - i k_2 l_2} \\ + \alpha^2 \omega_2
  \sum_{k_1,k_2=1}^{\infty} -i k_2 s_{k_1 k_2} e^{-i k_1 l_1 - i k_2 l_2} =
  \Ham^*_2\left( q_i^*, p_i^* \right),
\end{multline}
where
\begin{equation}
  S_2 = s_0 + \sum_{k_1,k_2 = 1}^{\infty} s_{k_1k_2} e^{-i k_1 l_1 - i
  k_2 l_2}.
\end{equation}
The terms dependent on $l_1$ will be eliminated from $\Ham^*_2$ if
\begin{equation}
  s_{k_1 k_2} = -i \frac{h_{k_1 k_2}}{\omega_1 k_1 + \omega_2 k_2}.
\end{equation}
Assuming than the system is far from resonance (that is, that
$\omega_1 k_1 + \omega_2 k_2 \neq 0$ for all $k_1$ and $k_2$), this
gives us the necessary $S_2$ to eliminate all terms in $\Ham_2$ that
depend on $l_1$ or $l_2$, leaving
\begin{equation}
  \Ham^*_2\left( q_i^*, p_i^* \right) =  \alpha^2 h_0 =
  \frac{1}{4\pi^2} \int_{0}^{2\pi} d l_1^* dl_2^*\, \Ham_2\left( q_i^*,
    p_i^* \right) .
\end{equation}
That is, our canonical transformation to eliminate the
rapidly-oscillating parts of $\Ham$ has left us with a Hamiltonian
that is the average over the oscillation period of the original
Hamiltonian\footnote{Note that the canonical variables are also
  transformed.  They differ from the original variables at
  $\mathcal{O}\left( \alpha^2 \right)$.  However, this difference is
  irrelevant when evaluating the interaction between the orbits
  described by $\Ham_2$, as this interaction is already
  $\mathcal{O}\left( \alpha^2 \right)$, and so the differences between
  the original and transformed variables contribute at sub-leading
  order.}.

The value of the Hamiltonian in equation~\eqref{eq:Ham4} averaged over
the mean motions is
\begin{eqnarray} %\label{eq:hami4}
\Ham^*_2&=&\frac{C_2}{8}\{  [1+3\cos(2i_2)] \big( [2+3e_1^2] [1+3\cos(2i_1)] \\ \nonumber 
&+&30e_1^2\cos(2g_1)\sin^2(i_1) \big) + 3\cos(2\Delta h) [10 e_1^2\cos(2g_1) \\ \nonumber
&\times&(3+\cos(2i_1)) + 4(2+3e_1^2)\sin(i_1)^2 ] \sin^2(i_2) \\ \nonumber
&+&12 ( 2+3e_1^2-5e_1^2\cos(2g_1))\cos(\Delta h)\sin(2i_1)\sin(2i_2)  \\ \nonumber
&+&120e^2_1\sin(i_1)\sin(2i_2)\sin(2g_1)\sin(\Delta h) \\ \nonumber
&-& 120e_1^2\cos(i_1)\sin^2(i_2)\sin(2g_1)\sin(2\Delta h) \} \ , 
\end{eqnarray}
where  $C_2$ was defied in equation (\ref{eq:C2}).

\subsection{  The Quadrupole--level Equations of Motion}\label{sec:quad_eq_motion}

% {\sn {I have created this a sub appendix of the quadrupole appendix
% here we find the equation of motions for the quadrupole approx }}

We use the canonical relations [equations (\ref{eq:Canoni})] in order
to derive the equations of motion from the Hamiltonian.  In our
treatment, both $H_1$ and $H_2$ evolve with time because the
Hamiltonian is not independent of $h_1$ and $h_2$.  From
eq.~(\ref{eq:H1}), we see that
\begin{equation}
\label{eq:Hdot1}
\dot{H}_1 = \frac{G_1}{G_{\tot}} \dot{G}_1 - \frac{G_2}{G_{\tot}} \dot{G}_2 \ ,
\end{equation}
and from eq.~\eqref{eq:z-sum-to-total} we see that
$\dot{H}_1=-\dot{H}_2$. The quadrupole-level Hamiltonian does not
depend on $g_2$; thus the magnitude of the outer orbit's angular
momentum, $G_2$, is constant\footnote{ This conserved quantity is lost
  at higher orders of the approximation; see \S \ref{sec:H8} and
  Appendix~\ref{sec:8}.}, and therefore
\begin{equation}
\label{eq:Hdot}
\dot{H}_1=\frac{G_1\dot{G}_1}{G_\tot} \ .
\end{equation}
From relations~(\ref{eq:Canoni}-\ref{eq:Canoni3}) we have
$\dot{H}_1=\partial \Ham / \partial h_1$, and $\dot{G}_1=\partial \Ham
/ \partial g_1$. The former gives
\begin{equation}
\label{eq:HdotsolF}
\dot{H}_1= -30 C_2 e_1^2\sin i_2 \sin i_\tot \sin (2g_1) \ .
\end{equation}
and the latter evaluates to
\begin{equation}
\label{eq:GdotsolF}
\dot{G}_1= -30 C_2 e_1^2 \sin^2 i_\tot \sin (2g_1) \ .
\end{equation}
Employing the law of sines, $G_\tot/ \sin i_\tot = G_1/ \sin i_2 = G_2
/ \sin i_1$, equation~\eqref{eq:HdotsolF} can also be written as
\begin{equation}
\label{eq:HdotsolF2}
\dot{H}_1= -\frac{G_1}{G_\tot} 30 C_2 e_1^2 \sin^2 i_\tot \sin (2g_1) \ ,
\end{equation}
which satisfies the relation in eq.~(\ref{eq:Hdot}).  The evolution of
the arguments of periapse are given by
\begin{eqnarray}
\label{eq:g1dot}
\dot{g}_1&=&6 C_2\bigg\{ \frac{1}{G_1}[4 \cos^2 i_\tot+(5\cos (2g_1)-1) \\ \nonumber
&\times& (1-e_1^2-\cos^2 i_\tot)] +\frac{\cos i_\tot}{G_2}[2 + e^2_1(3-5\cos (2g_1))]\bigg\} \ , %\\ \nonumber
\end{eqnarray}
and
\begin{eqnarray}
\label{eq:g2dot}
\dot{g}_2&=&3 C_2\bigg\{ \frac{2\cos i_\tot}{G_1}[2+e_1^2(3-5\cos (2g_1))] \\ \nonumber
&+&\frac{1}{G_2}[4+6e_1^2+(5\cos^2 i_\tot -3)(2+ e_1^2[3-5\cos(2g_1)] ) ] \bigg\} \ . %\\ \nonumber
\end{eqnarray}
Previous quadrupole-level calculations that made the substitution
error in the Hamiltonian lack the $1/G_2$ terms in these equations.
The evolution of the longitudes of ascending nodes is given by
\begin{equation}
\label{eq:h1dot_b}
\dot{h}_1=-\frac{3 C_2}{G_1 \sin i_1} \{ 2 + 3 e_1^2 - 5 e_1^2 \cos\left( 2 g_1 \right) \} \sin\left( 2 i_{\rm tot} \right)
\end{equation}
and
\begin{equation}
\label{eq:h1dot}
\dot{h}_2=-\frac{3 C_2}{G_2 \sin i_2} \{ 2 + 3 e_1^2 - 5 e_1^2 \cos\left( 2 g_1 \right) \} \sin\left( 2 i_{\rm tot} \right).
\end{equation}
Using the law of sines, $G_1 \sin i_1 = G_2 \sin i_2$, from which we
get $\dot{h}_1 = \dot{h}_2$, as required by the relation $h_1 - h_2 =
\pi$.  In many systems it is useful to calculate the time evolution of
the eccentricity, obtained through the following relation:
\begin{equation}
\frac{de_j}{dt}=\frac{\partial e_j}{\partial G_j}\frac{\partial \Ham}{\partial g_j} \ ,
\end{equation}
%which can also be written as follows:
%\begin{equation}
% \frac{de_j}{dt}= \frac{dG_j}{dt} \left( \frac{\partial G_j}{\partial e_j} \right)^{-1} = \frac{1-e_j2}{e_j G_j} \frac{d G_j}{dt}.
%\end{equation}
In the quadrupole approximation $\dot{e}_2=\dot{G}_2 = 0$ (which is
not the case at higher order in $\alpha$; see Appendix
\ref{sec:8}). The eccentricity evolution for the inner orbit is given
by
\begin{equation}
\dot{e}_1=C_2\frac{1-e_1^2}{G_1}30e_1 \sin^2i_\tot\sin(2g_1) \ .
\end{equation}
Another useful parameter is the inclination, which can be found
through the $z$-component of the angular momentum:
\begin{equation}
\frac{d (\cos i_1)}{dt}= \frac{\dot{H}_1}{G_1}-\frac{\dot{G}_1}{G_1} \cos i_1 \ ,
\end{equation}
and similarly for $i_2$ (but note again that $\dot{G}_2 = 0$ to
quadrupole order).

%%%%%%%%%%%%%%%%%%%%%%%%%%%%%%%%%%%%%%%%%%%%%%%%%%%%%%%%%%%%%%
%%%%%%%%%%%%%%%%%%%%%%%%%%%%%%%
\subsection{ Maximum Eccentricity and ``Kozai'' Angles in the Quadrupole Approximation}\label{sec:maxmin}

First note that setting $\dot{e_1}=0$ also means that $\dot{G}_1=0$.
The values of the argument of periapsis that satisfy these relations
are: $g_1=0+\pi n/2$, where $n=0,1,2...$ .  Also, setting
$\dot{G}_{1}(e_{1,{\rm max,min}})=0$ means that $\dot{H}_{1}(e_{1,{\rm
    max,min}})=0$ and $\dot{i}_{1}=0$, i.e., an extremum of the
eccentricity is also an extremum of both the inner and outer
inclinations.
 
The conservation of the total angular momentum, i.e., ${\bf G}_1+{\bf
  G}_2={\bf G}_\tot$ sets the relation between the total inclination
and inner orbit eccentricity. We re-write equation (\ref{eq:cosi}) as
\begin{equation}
 \label{eq:ang}
L_1^2(1-e_1^2)+2L_1L_2\sqrt{1-e_1^2}\sqrt{1-e_2^2}\cos i_\tot = G_\tot^2-G_2^2 \ ,
\end{equation}
where in the quadrupole-level approximation $e_2$ and $G_2$ are
constant.  The right hand side of the above equation is set by the
initial conditions. In addition, $L_1$, and $L_2$ [see
eqs.~(\ref{eq:L1}) and (\ref{eq:L2})] are also set by the initial
conditions.  Using the conservation of energy we can write, for the
minimum eccentricity case (i.e., setting $g_1=0$)
 \begin{equation}
 \label{eq:Ep}
\frac{E}{2 C_2}=3\cos ^2 i_\tot (1-e_1^2)-1+6e_1^2 \ ,
\end{equation}
where we also used the relation $\Delta h=\pi$. We find a similar
equation if we set $g_1=\pi/2$:
 \begin{equation}
 \label{eq:Em}
\frac{E}{2 C_2}=3\cos ^2 i_\tot (1+4e_1^2)-1-9e_1^2 \ .
\end{equation}
Equations (\ref{eq:ang}), (\ref{eq:Ep}) and (\ref{eq:Em}) give a
simple relation between the total inclination and the inner
eccentricity.  The remainder of the parameters in the equations are
defined by the initial conditions.  Thus, using equations
(\ref{eq:Ep}) and (\ref{eq:ang}) we can find the minimum eccentricity
reached during the oscillation and using equations (\ref{eq:Em}) and
(\ref{eq:ang}) we can find also the maximum and the minimum
inclinations.  The following example illustrates the relation defined
by these equations between the inclination and the eccentricity.

\begin{figure}
\begin{center}
%\plotone{maxmin4.eps}
\includegraphics[width=84mm]{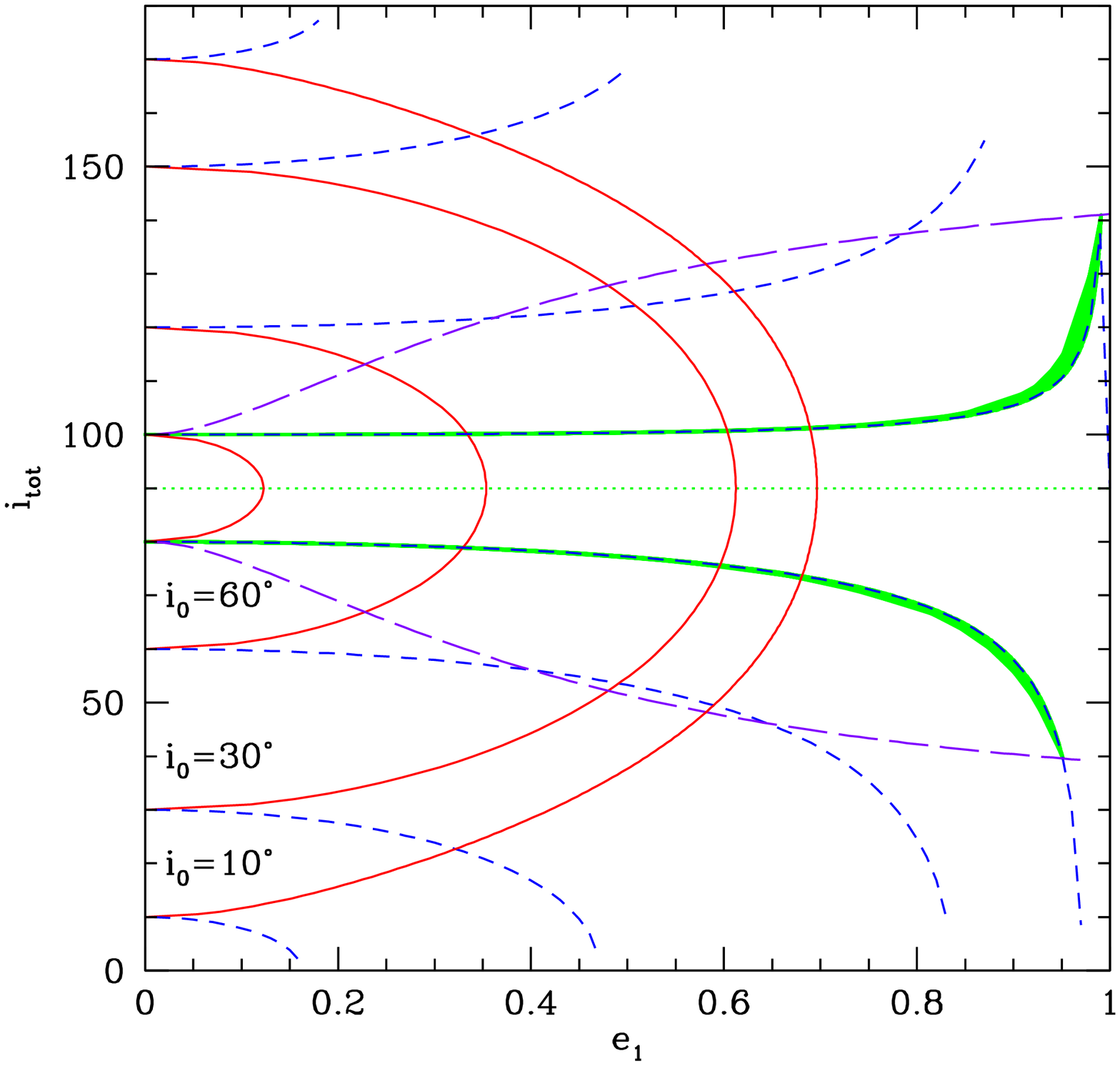}
\caption{The total inclination and eccentricity relation for an Algol--like system. We show
  constant energy curves (solid curves, Eq.~\eqref{eq:Ep_in}) and
  constant total angular momentum curves (dashed curves,
  Eq.~\eqref{eq:ang_in}). The initial conditions considered here are
  $e_1^0=0$, $g_1^0$, $e_2^0=0$ and $L_1/L_2=0.07$, appropriate for
  the Algol system (see Section~\ref{sec:3S}).  We consider four
  different initial inclinations and their symmetric $90^\circ$
  counterparts, from bottom to top $10,30,60$ and $80$ degrees. We
  also show an example (highlighted curve) for the  system which
  is a result of integration of the quadrupole-level approximation
  equations. } \label{fig:maxmin4}
\end{center}
\end{figure}

For simplicity we set initially $e_1^0=0$, $g_1^0$ and $e_2^0=0$ (the
superscript $0$ stand for initial values). In this appendix we
consider only the quadrupole-level approximation, and thus $e_2$
doesn't change. Using these initial conditions (and for some initial
mutual inclination $i_0$) we can write equation (\ref{eq:ang}) as
 \begin{equation}
 \label{eq:ang_in}
 \sqrt{1-e^2_1} \cos i_\tot=\cos i_0+\frac{L_1}{2 L_2}e_1^2  \ .
\end{equation}
We show these curves for different $i_0$ in Figure \ref{fig:maxmin4}
(short dashed curves) for a hypothetical system with the parameters of
 an Algol--like system (but with $e_2=0$, see \S \ref{sec:3S}). Note that there is a
slight asymmetry between the prograde and retrograde orbits due to the
$L_1/L_2$ factor \citep[which is not the case for the test particle
case, see][]{LN,Boaz2}.  Similar analysis for the Algol system was done in \citet[][Figure 1]{sod06}. We also write equations (\ref{eq:Ep}) and
(\ref{eq:Em}) using the initial conditions.  Equation (\ref{eq:Ep})
can be simplified to
\begin{equation}
 \label{eq:Ep_in}
(1-e^2_1)\cos^2i_\tot = \cos^2 i_0-2e_1^2 \ ,
\end{equation}
depicted in Figure \ref{fig:maxmin4} (solid curves, for different
$i_0$). As can be seen from the Figure, this equation gives the
minimum eccentricity, which is the crossing point with equation
(\ref{eq:ang_in}). For these choice of initial conditions the minimum
eccentricity is $e_1^0 = 0$.  Equation (\ref{eq:Em}) becomes
\begin{equation}
 \label{eq:Em_in}
(1+4e^2_1)\cos^2i_\tot = \cos^2 i_0+3e_1^2 \ ,
\end{equation}
which is depicted in Figure \ref{fig:maxmin4} (long dashed curves, for
$i_0=80^\circ$ and $100^\circ$).  We now use this equation and
equation (\ref{eq:ang_in}) to find the maximum eccentricity. After
some algebra we find:
\begin{eqnarray}
\left(\frac{L_1}{L_2}\right)^2 e_1^4 + \left(3+4\frac{L_1}{L_2}\cos i_0 + \left(\frac{L_1}{2L_2}\right)^2 \right)e_1^2 \nonumber \\
+\frac{L_1}{L_2}\cos i_0-3+5\cos^2 i_0 = 0 \ .
\end{eqnarray}
As we approach the TPQ limit, $L_2 \gg L_1$, and this equation becomes
\begin{equation}
\label{eq:maxe}
e^2_1=1-\frac{5}{3}\cos^2 i_0 \ ,
\end{equation}
which gives the maximum eccentricity as a function of mutual initial
inclination with zero initial inner eccentricity.  In Figure
\ref{fig:maxmin4} we show that this approximation still holds fairly
well even for  an Algol--like system, where $L_1/L_2 \sim 0.07$.  Equation
(\ref{eq:maxe}) has been found previously
\citep[e.g.][]{Inn+97,Kin+99,3book} in the TPQ approximation, but in
these works it is assumed valid outside that limit.  A solution exists
only if the right hand side of this equations is positive, thus we
find the critical angles for large Kozai oscillation in the TPQ limit:
\begin{equation}
39.2^\circ \leq i_0 \leq 140.8^\circ \ .
\end{equation}
For larger $L_1/L_2$ and/or for initial $e_1>0$ this limit and $e_{\rm
  max}$ are different and the full solution of equations
\eqref{eq:ang},\eqref{eq:Ep} and \eqref{eq:Em} is required. In fact
for each initial set of $e_1>0$ and $i_\tot$, there is a specific
$L_1/L_2$ that will produce an angular momentum curve that crosses
$90^\circ$. Thus, for initial $g_1>90^\circ$ the mutual inclination
can oscillate from value below $90^\circ$ to above. This happens
because the inclination of the outer orbit $i_2$ changes considerably,
while the inner orbit retains its prograde or retrograde orientation.

%%%%%%%%%%%%%%%%%%%%%%%%%%%%%%%
\section{ The Full Octupole-Order  Equations of Motion}\label{sec:8}

We define:
\begin{equation}
\label{eq:C3}
C_3=-\frac{15}{16}\frac{k^4}{4}\frac{(m_1+m_2)^9}{(m_1+m_2+m_3)^4}\frac{m_3^9(m_1-m_2)}{(m_1m_2)^5}\frac{L_1^6}{L_2^3G_2^5} \ .
\end{equation}
Note that this definition differs in sign sign from \citet{Ford00},
and is consistent with \citet{Bla+02,Ford04}. For $m_1 = m_2$ this
factor is zero. We also define:
\begin{equation}
\label{eq:A8}
A= 4+3e_1^2-\frac{5}{2}B\sin i_\tot^2 \ ,
\end{equation}
where
\begin{equation}
B=2+5e^2_1-7e_1^2\cos(2g_1) \ ,
\end{equation}
and
\begin{equation}
\cos \phi=-\cos g_1\cos g_2 -\cos i_\tot \sin g_1 \sin g_2 \ .
\end{equation}
%C_3e_2\bigg\{e_1 \left(\frac{1}{G_2}+\frac{\cos i_\tot}{G_1}\right)
%
 
As mentioned in Section \ref{sec:H8} the evolution equations for $e_2,
g_2, g_1$ and $e_1$ can be found correctly from a Hamiltonian that has
had $h_1$ and $h_2$ eliminated by the relation $h_1 - h_2 = \pi$; the
partial derivatives with respect to the other coordinates and momenta
are not affected by the substitution. The time evolution of $H_1$ and
$H_2$ (and thus $i_1$ and $i_2$) can be derived from the total angular
momentum conservation. Thus it is useful to write the much simpler the
doubly averaged Hamiltonian after eliminating the nodes:
\begin{eqnarray}
  \Ham(\Delta h \to \pi)& =& C_2 \{ \left( 2 + 3 e_1^2 \right) \left( 3 \cos^2 i_\tot - 1 \right)  \\ & + &15 e_1^2 \sin^2 i_\tot \cos(2 g_1) \}  \nonumber \\ \nonumber
  &+& C_3 e_1 e_2 \{ A \cos \phi \nonumber \\ &+ & 10 \cos i_\tot \sin^2 i_\tot  (1-e_1^2) \sin g_1 \sin g2 \} \nonumber \ .
\end{eqnarray} 

The time evolution of the argument of periapse for the inner and outer
orbits are given by:
\begin{eqnarray}
\label{eq:g1dot8} 
\dot{g}_1&=&6 C_2\bigg\{ \frac{1}{G_1}[4 \cos^2 i_\tot+(5\cos (2g_1)-1)  \\ \nonumber 
&\times& (1-e_1^2-\cos^2 i_\tot)] +\frac{\cos i_\tot}{G_2}[2 + e^2_1(3-5\cos (2g_1))]\bigg\} \\ \nonumber
&-&C_3e_2\bigg\{e_1 \left(\frac{1}{G_2}+\frac{\cos i_\tot}{G_1}\right) \\ \nonumber
&\times& [\sin g_1 \sin g_2(10(3\cos^2 i_\tot-1)(1-e_1^2)+A) \\ \nonumber
&-&5 B \cos i_\tot\cos \phi] -\frac{1-e_1^2}{e_1G_1}\times[\sin g_1\sin g_2 \\ \nonumber
&\times&10\cos i_\tot \sin i_\tot^2(1-3e_1^2) \\ \nonumber
&+&\cos\phi (3A-10\cos i_\tot^2+2)]\bigg\} \ ,
\end{eqnarray}
and 
\begin{eqnarray}
\label{eq:g2dot8}
\dot{g}_2&=&3 C_2\bigg\{ \frac{2\cos i_\tot}{G_1}[2+e_1^2(3-5\cos (2g_1))]  \\ \nonumber 
&+&\frac{1}{G_2}[4+6e_1^2+(5\cos^2 i_\tot -3)(2+ e_1^2[3-5\cos(2g_1)] )\bigg\} \\ \nonumber
&+&C_3e_1  \bigg\{\sin g_1\sin g_2 \bigg ( \frac{4e_2^2+1}{e_2G_2} 10\cos i_\tot\sin^2 i_\tot (1-e_1^2) \\ \nonumber
&-&e_2\left(\frac{1}{G_1}+\frac{\cos i_\tot}{G_2}\right) [A+10(3\cos^2 i_\tot-1)(1-e^2_1)]\bigg ) \\ \nonumber
&+&\cos \phi \bigg [ 5B\cos i_\tot e_2 \left(\frac{1}{G_1}+\frac{\cos i_\tot}{G_2}\right) +\frac{4e_2^2+1}{e_2G_2}A\bigg ]\bigg \}
\end{eqnarray}
The time evolution of the longitude of ascending nodes is given by:
\begin{eqnarray}
\label{eq:h1dot8}
\dot{h}_1&=&-\frac{3 C_2}{G_1 \sin i_1} \left( 2 + 3 e_1^2 - 5 e_1^2 \cos\left( 2 g_1 \right) \right) \sin\left( 2 i_{\rm tot} \right) \\ \nonumber
&-&C_3e_1e_2[5B\cos i_\tot\cos\phi \nonumber \\ &-& A\sin g_1\sin g_2 +10(1-3\cos^2 i_\tot) \nonumber \\ \nonumber
&\times&(1-e_1^2)\sin g_1\sin g_2]\frac{\sin i_\tot}{G_1 \sin i_1} \ ,
\end{eqnarray}
where in the last part we have used again the law of sines for which
$\sin i_1= G_2 \sin i_\tot / G_\tot$.  The evolution of the longitude
of ascending nodes for the outer orbit can be easily obtained using:
\begin{equation}
\dot{h}_2=\dot{h}_1 \ .
\end{equation}

The evolution of the eccentricities is:
\begin{eqnarray}
\label{eq:dote1_8}
\dot{e}_1&=&C_2\frac{1-e_1^2}{G_1}[30e_1 \sin^2i_\tot\sin(2g_1)] \\ \nonumber
&+& C_3e_2\frac{1-e_1^2}{G_1}[35 \cos\phi\sin^2 i_\tot e_1^2 \sin (2g_1) \\ \nonumber
&-& 10\cos i_\tot \sin^2 i_\tot \cos g_1 \sin g_2 (1-e_1^2)\\ \nonumber
&-& A(\sin g_1\cos g_2  - \cos i_\tot \cos g_1 \sin g_2)] \ ,
\end{eqnarray}
and
\begin{eqnarray} 
\dot{e}_2 &=& -C_3 e_1\frac{1-e_2^2}{G_2}[10\cos \left(i_\tot\right) \sin^2 \left(i_\tot\right)(1-e_1^2)\sin g_1 \cos g_2 \nonumber \\ 
&+& A(\cos g_1 \sin g_2 - \cos(i_\tot)\sin g_1 \cos g_2)] \ .
\end{eqnarray}

We also write the angular momenta derivatives as a function of time;
for the inner orbit
\begin{eqnarray} 
\label{eq:G1dot8}
\dot{G}_1 & =& - C_2 30 e_1^2  \sin (2g_1) \sin^2(i_\tot) + C_3 e_1 e_2 ( \\ \nonumber
&-&35 e_1^2 \sin^2(i_\tot)\sin (2 g_1) \cos\phi + A [\sin g_1 \cos g_2  \\ \nonumber
&-& \cos(i_\tot)\cos g_1 \sin g_2 ]  \\ \nonumber
&+& 10\cos(i_\tot)\sin^2(i_\tot)
[1-e_1^2]\cos g_1 \sin g_2  )\ ,
\end{eqnarray}
and for the outer orbit (where the quadrupole term is zero)
\begin{eqnarray} 
\label{eq:G2dot8}
\dot{G}_2 & =&  C_3 e_1 e_2 [A\{\cos g_1 \sin g_2  - \cos(i_\tot)\sin
g_1 \cos g_2 \}\nonumber \\  &+&10\cos(i_\tot)\sin^2(i_\tot)
[1-e_1^2]\sin g_1 \cos g_2  ]\ .
\end{eqnarray}
Also,
\begin{equation}
\label{eq:Hdot1_8}
  \dot{H}_1 = \frac{G_1}{G_{\tot}} \dot{G}_1 - \frac{G_2}{G_{\tot}} \dot{G}_2 \ ,
\end{equation} 
where using the law of sines we write:
\begin{equation}
\label{eq:Hdot1_8n}
  \dot{H}_1 = \frac{\sin i_2}{\sin i_{\tot}} \dot{G}_1 - \frac{\sin i_1}{\sin i_{\tot}} \dot{G}_2 \ .
\end{equation} 
The inclinations evolve according to
\begin{equation}
\dot{(\cos i_1)}= \frac{\dot{H}_1}{G_1}-\frac{\dot{G}_1}{G_1} \cos i_1 \ ,
\end{equation}
and
\begin{equation}
\dot{(\cos i_2)}= \frac{\dot{H}_2}{G_2}-\frac{\dot{G}_2}{G_2} \cos i_2 \ .
\end{equation}
Our equations are equivalent to those of \citet{Ford00}, but we give
the evolution equations for $H_1$ and $H_2$ (and $i_1$ and $i_2$).

%%%%%%%%%%%%%%%%%%%%%%%%%%%%%%%

\section{ Elimination of the Nodes and the Problem in Previous
  Quadrupole-Level Treatments }\label{prob}

Since the total angular momentum is conserved, the ascending nodes
relative to the invariable plane follow a simple relation, $h_1(t) =
h_2(t) -\pi$.  If one inserts this relation into the Hamiltonian,
which only depends on $h_1 - h_2$, the resulting ``simplified''
Hamiltonian is independent of $h_1$ and $h_2$.  One might be tempted
to conclude that the conjugate momenta $H_1$ and $H_2$ are constants
of the motion.  However, that conclusion is false.  This incorrect
argument has been made by a number of authors%
\footnote{ For example, \citet[p.\ 592]{Kozai} incorrectly argues that
  ``As the Hamiltonian $F$ depends on $h$ and $h'$ as a combination
  $h-h'$, the variables $h$ and $h'$ can be eliminated from $F$ by the
  relation (5). Therefore, $H$ and $H'$ are constant."}. %

In general, using \emph{dynamical} information about the system---in
this case that angular momentum is conserved, implying that ${\bf G}_1
+ {\bf G}_2 = {\bf G}_\tot$ at all times and therefore $h_1 - h_2 =
\pi$---to simplify the Hamiltonian is not correct.  The derivation of
Hamilton's equations relies on the possibility of making
\emph{arbitrary} variations of the system's trajectory, and such
simplifications restrict the allowed variations to those which respect
the dynamical constraints.  Once Hamilton's equations are employed to
derive equations of motion for the system, however, dynamical
information can be employed to simplify these equations.

In our particular case, equations of motion for components of the
system that do not involve partial derivatives with respect to $h_1$
or $h_2$ will not be affected by the node-elimination substitution.
For this reason, it is correct to derive equations of motion for all
components \emph{except} for $H_1$ and $H_2$ from the node-eliminated
Hamiltonian; expressions for $\dot{H}_1$ and $\dot{H}_2$ can then be
derived from conservation of angular momentum.  This approach has been
employed in at least one computer code for octupole evolution, though
the discussion in the corresponding paper incorrectly eliminates the
nodes in the Hamiltonian \citep{Ford00}.

In some later studies,
\citep{Sid83,Inn+97,1998KEM,1998EKH,Mik+98,Kin+99,Egg+01,Wu+03,3book,Dan,Wu+07,Zdz+07,PF09},
the assumption that $H_1={\rm const}$ (i.e.~the TPQ approximation) was
built into the calculations of quadrupole-level secular evolution for
various astrophysical systems, even when the condition $G_2 \gg G_1$
was not satisfied. Moreover many previous studies simply set
$i_2=0$. This is equivalent to the TPQ approximation; for non-test
particles, given the mutual inclination $i$, the inner and outer
inclinations $i_1$ and $i_2$ are set by the conservation of total
angular momentum [see equations (\ref{eq:i1}) and (\ref{eq:i2})].

%%%%%%%%%%%%%%%%%%%%%%%%%%%%%%%
\end{document}